\documentclass[12pt]{article}
\pdfoutput=1
\usepackage{setspace}

\usepackage{graphicx}
\usepackage{epstopdf}
\usepackage[body={18cm, 21cm},right=1.8cm]{geometry}
\usepackage{amssymb}
\usepackage{amsmath}
\usepackage{dcolumn}

\usepackage{graphicx}
\usepackage{epstopdf}
\usepackage{amsmath}
\usepackage{amsfonts}
\usepackage{amssymb}
\usepackage[usenames]{color}
\usepackage{mathrsfs}
\usepackage{array}
\usepackage{nonfloat}

\usepackage{multirow}
\usepackage{psfrag}
\pdfoutput=1

\usepackage[colorlinks,bookmarks]{hyperref}
\definecolor{linkblue}{rgb}{0,0,0.8}
\definecolor{linkgreen}{rgb}{0,0.5,0}

\hypersetup{pdfpagemode=UseNone, pdfstartview=FitH, linkcolor=linkblue, %
            citecolor=linkgreen, urlcolor=linkblue}

\newcommand\nn{\nonumber}
\newcommand\eea{\end{eqnarray}}
\newcommand\bea{\begin{eqnarray}}

\def\la{\langle}
\def\ra{\rangle}
\def\beq{\begin{equation}}
\def\eeq{\end{equation}}
\def\d{\partial}

\newcommand{\be}{\begin{equation}}
\newcommand{\ee}{\end{equation}}
\newcommand{\ba}{\begin{align}}
\newcommand{\ea}{\end{align}}
\newcommand{\bg}{\begin{gather}}
\newcommand{\eg}{\end{gather}}
\newcommand{\bseq}{\begin{subequations}}
\newcommand{\eseq}{\end{subequations}}

\newcommand{\vk}{\boldsymbol{k}}
\newcommand{\vkp}{\boldsymbol{q}}
\newcommand{\vkkp}{\boldsymbol{k}-\boldsymbol{q}}
\newcommand{\vq}{\boldsymbol{q}}

\newcommand{\vx}{\boldsymbol{x}}

\newcommand{\vv}{\boldsymbol{v}}

\def\lcdm{$\Lambda$CDM}
\newcommand{\knl}{k_{\rm NL}}

\def\H{{\cal H}}
\newcommand{\hinvMpc}{h\,$Mpc$^{-1}}
\newcommand{\ktr}{k_{\rm tr}}

\newcommand{\tknl}{{\tilde k}_{\rm NL}}

\newcommand{\invMpc}{\,h\, {\rm Mpc}^{-1}\,}

\def\Omm{\Omega_{\rm m}}

\newcommand{\lp}{\left(}
\newcommand{\rp}{\right)}
\newcommand{\lb}{\left[}
\newcommand{\rb}{\right]}
\def\co{c_{s  (1)}^2}

\definecolor{purple}{rgb}{0.78,0.18,0.77}

\newcommand{\gammai}{(\d\tau)_{\rho_l}}

\newcommand{\Mpc}{{\rm Mpc}}
\newcommand{\epsm}{\epsilon_{s_<}}
\newcommand{\epsp}{\epsilon_{s_>}}

\makeatletter
\newlength{\apb@width}
\newcommand{\autoparbox}[2][c]{\settowidth{\apb@width}{#2}\parbox[#1]{\apb@width}{#2}}

\makeatother

\begin{document}

\vspace{5mm}
\vspace{0.5cm}
\begin{center}

\def\thefootnote{\fnsymbol{footnote}}

{\Large \bf The One-Loop Matter Bispectrum in the\\[0.4cm]
Effective Field Theory of
Large Scale Structures
}
\\[0.8cm]

{\large Raul E.~Angulo$^{1}$, Simon Foreman$^{2,3}$, \\[0.4cm]
 Marcel Schmittfull$^{4}$, and Leonardo Senatore$^{2,3}$}
\\[0.5cm]

{\normalsize { \sl $^{1}$ Centro de Estudios de Fisica del Cosmos de Aragon,\\ Plaza San Juan 1, Planta-2, 44001, Teruel, Spain}}\\
\vspace{.3cm}

{\normalsize { \sl $^{2}$ Stanford Institute for Theoretical Physics and Department of Physics, \\Stanford University, Stanford, CA 94306}}\\
\vspace{.3cm}

{\normalsize { \sl $^{3}$ Kavli Institute for Particle Astrophysics and Cosmology, \\ SLAC  and Stanford University, Menlo Park, CA 94025}}\\
\vspace{.3cm}

{\normalsize { \sl $^{4}$Berkeley Center for Cosmological Physics, Department of Physics,\\
University of California Berkeley and Lawrence Berkeley National
Laboratory, Berkeley, CA 94720 }}\\
\vspace{.3cm}

\end{center}

\vspace{.8cm}

\hrule \vspace{0.3cm}
{\small  \noindent \textbf{Abstract} \\[0.3cm]
\noindent  Given the importance of future large scale structure surveys for delivering new cosmological information, it is crucial to reliably predict their observables.  The Effective Field Theory of Large Scale Structures (EFTofLSS) provides a manifestly convergent perturbative scheme to compute the clustering of dark matter in the weakly nonlinear regime in an expansion in $k/k_{\rm NL}$, where $k$ is the wavenumber of interest and $k_{\rm NL}$ is the wavenumber associated to the nonlinear scale. It has been recently shown that the EFTofLSS matches to $1\%$ level the dark matter power spectrum at redshift zero up to $k\simeq 0.3 h\,$Mpc$^{-1}$ and $k\simeq 0.6 h\,$Mpc$^{-1}$ at one and two loops respectively, using only one counterterm that is fit to data. Similar results have been obtained for the momentum power spectrum at one loop. This is a remarkable improvement with respect to former analytical techniques. Here we study the prediction for the equal-time dark matter bispectrum at one loop. We find that at this order it is sufficient to consider the same counterterm that was measured in the power spectrum. Without any remaining free parameter, and in a cosmology for which $k_{\rm NL}$ is smaller than in the previously considered cases ($\sigma_8=0.9$), we find that the prediction from the EFTofLSS  agrees very well with $N$-body simulations up to $k\simeq 0.25 h\,$Mpc$^{-1}$, given the accuracy of the measurements, which is of order a few percent at the highest $k$'s of interest. While the fit is very good on average up to  $k\simeq 0.25\hinvMpc$, the fit performs slightly worse on equilateral configurations, in agreement with expectations that for a given maximum $k$, equilateral triangles are the most nonlinear.
\noindent 
}

 \vspace{0.3cm}
\hrule
\def\thefootnote{\arabic{footnote}}
\setcounter{footnote}{0}

\vspace{.8cm}

\newpage
\tableofcontents

\section{Introduction}

To explore the dynamics that drove the inflationary period, it is necessary to probe a large number of modes. The upcoming release of the polarization data from the Planck satellite will increase the amount of available data by at most a factor of two, which means that our knowledge of inflation will be improved by a similar amount, depending on the quantity being constrained. After this, no significant increase in the number of available modes is expected from the CMB, even though some high-$\ell$ measurement might still contain some interesting information. Notice that this situation applies quite unaltered also in the case in which the recent claimed discovery of primordial $B$ modes from the BICEP experiment~\cite{Ade:2014xna} is confirmed. Such a measurement would indeed imply a discovery that Inflation happened at high energies, but it would not enlighten us in detail on the dynamics that drove inflation, for which we need more modes. 

This tells us that it is crucial for the field of cosmology to find ways of accessing more primordial modes.
The best option to find more modes in the next decade is from large scale structure (LSS) surveys. It is still unclear how many modes will contain relevant cosmological information, and in particular how many of these modes will be accessible with analytical techniques. In order to address this question, the Effective Field Theory of Large Scale Structures (EFTofLSS) was developed in~\cite{Carrasco:2012cv,Baumann:2010tm}. The EFTofLSS is different from all former available approaches, such as SPT~\cite{Bernardeau:2001qr} or RPT~\cite{Crocce:2005xy}, for the fact that it acknowledges that analytical predictions cannot be made to arbitrarily small scales where the dynamics is completely nonlinear. The effect of short scales on long modes is encapsulated in a series of terms in the equations of motion of the long modes, whose numerical coefficients are not known, and need to be fitted to data or to $N$-body simulations.
The role of these additional terms is to compensate for any erroneous dependence on short distance physics that appears when performing calculations in perturbation theory. This is what, in the context of particle physics, goes under the name of renormalization, which is why these additional terms that appear in the equations of motion are often referred to as counterterms.

As described in~\cite{Porto:2013qua,Senatore:2014via}, after the inclusion of these counterterms and after renormalization, perturbation theory amounts to an expansion in the parameters: $\epsilon_{\delta<}$ and $\epsilon_{s>}$. These are defined as 
\bea
&&\epsilon_{s >} =k^2  \int_k^\infty {d^3k' \over (2 \pi)^3}  {P_{11}(k') \over k'^2}\ , \\ \nonumber
&&\epsilon_{\delta <} = \int^k_0 {d^3k' \over (2 \pi)^3} P_{11}(k')\ ,
\eea
where $P_{11}(k)$ is the linear dark matter power spectrum. $\epsilon_{\delta <}$ represents the effect of tidal forces on a given region, while $\epsilon_{s >}$ accounts for the short distance displacements. Both of these scale  proportionally to powers of $k/\knl$, where $\knl$ is a wavenumber related to the nonlinear scale~\footnote{We give a precise definition of $\knl$ later in the text. The place where the EFTofLSS ultimately fails is  related to~$\knl$ by order one factors that cannot be reliably estimated without detailed calculations. This scale might actually be considered as the true nonlinear scale of the universe, but, as typical for scales where the theory is strongly coupled, it is impossible to determine them without order one ambiguity unless one performs very high order calculations.}. In the most common treatment, the so-called Eulerian one, we expand also in the effect of the long distance displacements $\epsilon_{s<}= (k\, \delta s_<)^2$, where
\bea
\epsm &=&k^2  \int_0^k {d^3k' \over (2 \pi)^3}  {P_{11}(k') \over k'^2}\ .
\eea
As described in~\cite{Senatore:2014via}, it is not possible in general to expand in $\epsm$, as  for the modes of interest this parameter is of order one in our universe. However, there are some quantities, so called IR-safe, for which the effect of $\epsm$ almost cancels out completely, leaving only a small $2\%$ effect connected to the Baryon Acoustic Oscillations. In general, to deal with the effects related to $\epsm$, a resummation of the IR-modes needs to be performed, as it was done in~\cite{Senatore:2014via}. After this is done, the expansion parameters in the EFTofLSS remain only $\epsp$ and $\epsilon_{\delta <} $, which are smaller than one for $k\lesssim \knl$.  This means that, apart for non-perturbative effects that dominate at very small distances, perturbation theory with the EFTofLSS is manifestly convergent.

So far, the EFTofLSS has been used to predict the power spectrum of dark matter at redshift zero. The results have been incredibly encouraging. The results at two loops~\cite{Carrasco:2013mua}, after IR resummation~\cite{Senatore:2014via}, agree at 1\% with $N$-body simulations to the remarkably high wavenumber of $k\simeq 0.6\hinvMpc$. This represents an improvement of about a factor of six in wavenumber with respect to former analytic techniques, such as SPT.
Since available modes scale as the cube of the maximum wavenumber that can be predicted, $k_{\rm max}^3$, these results tell us that there is potentially a factor of about 200 more modes that are amenable to an analytic treatment in next generation LSS surveys than previously believed~\footnote{Of course this is a na\"{i}ve extrapolation of the results obtained at $z=0$. A more careful estimate would require performing the same study at all redshifts, that we defer to an upcoming paper~\cite{Foreman:2015uva}.}. Such a potentially picture-changing result makes it very important to check if a comparable improvement persists for every observable. In~\cite{Senatore:2014via}, the momentum power spectrum, which is not an IR-safe quantity, was computed at one loop, where it was shown that, without the need of fitting any additional parameter, the prediction for the momentum showed a gain in the UV reach that was comparable to the one seen for the matter power spectrum, with percent agreement up to $k\simeq 0.3\hinvMpc$. 
At this point it is tempting to see if a similar improvement holds for higher $n$-point functions. The easiest one is the bispectrum at one-loop, which is the subject of study of this paper.

\section{Formulas for One-Loop EFT Prediction}

We wish to obtain a prediction for the matter bispectrum $B(k_1,k_2,k_3)$, defined by
\beq
\left\la \delta(\vk_1) \delta(\vk_2) \delta(\vk_3) \right\ra
	= (2\pi)^3 \delta_{\rm D}(\vk_1+\vk_2+\vk_3) B(k_1,k_2,k_3)\ ,
\eeq
where the Dirac delta function~$\delta_{\rm D}$ is written with a subscript to distinguish it from the matter overdensity~$\delta$. Note that, by translation and rotation invariance, the bispectrum is a function of the magnitudes of three wavevectors that add to form a triangle in momentum space, and does not depend on the specific orientation of the triangle. The bispectrum also depends on the time of observation, but in this work we are only concerned with the predictions at $z=0$, so we will suppress the time-dependence in what follows.

App.~\ref{sec:eftreview} reviews the generic method for solving the equations of motion for $\delta$ and the velocity field~$v^i$ and thereby deriving expressions for their correlation functions, and also defines the relevant notation for the solutions (most of which will be familiar from the literature involving SPT, e.g.~\cite{Bernardeau:2001qr}). Using the procedure outlined in this appendix, and also described in~\cite{Carrasco:2013mua} (see also~\cite{Carrasco:2012cv}), the lowest-order expression for the bispectrum, which we call $B_\text{tree}$ because it is a tree graph if written out diagrammatically, is found to be
\beq
\label{eq:btree}
B_{\rm tree}(k_1,k_2,k_3) = 
	2P_{11}(k_1) P_{11}(k_2) F_2^{\rm (s)}(\vk_1,\vk_2) + \text{2 permutations}\ ,
\eeq
where $P_{11}(k)$ is the linear matter power spectrum. The next-order corrections to this expression are the one-loop SPT diagrams $B_\text{1-loop}$, discussed in Sec.~\ref{sec:irsafeintegrand}, and various counterterms arising from the terms in the effective stress tensor, which have not previously been included in bispectrum calculations and which we discuss in Sec.~\ref{sec:counterterms}.

\subsection{The IR-Safe Integrand of $B_\text{1-loop}$}
\label{sec:irsafeintegrand}

The one-loop correction in SPT is commonly written as the sum of four terms (e.g.~\cite{Scoccimarro:1997st}):
\beq
B_\text{1-loop} = B_{222} + B_{321}^{\rm (I)} + B_{321}^{\rm (II)} + B_{411}\ ,
\eeq
where
\begin{align}
\nn
B_{222}(k_1,k_2,k_3) &= 8 \int_{\vq} P_{11}(q) P_{11}(|\vk_2-\vq|) P_{11}(|\vk_3+\vq|) \\
&\qquad \times F_2^{\rm (s)}(-\vq,\vk_3+\vq)
	F_2^{\rm (s)}(\vk_3+\vq,\vk_2-\vq) F_2^{\rm (s)}(\vk_2-\vq,\vq)\ , 
	\label{eq:b222scoc} \\
\nn
B_{321}^{\rm (I)}(k_1,k_2,k_3) &= 6P_{11}(k_3) \int_{\vq} P_{11}(q) P_{11}(|\vk_2-\vq|) \\
&\qquad \times F_3^{\rm (s)}(-\vq,-\vk_2+\vq,-\vk_3) F_2^{\rm (s)}(\vq,\vk_2-\vq)
	+ \text{5 permutations}\ ,
	\label{eq:b321onescoc} \\
B_{321}^{\rm (II)}(k_1,k_2,k_3) &= 6P_{11}(k_2) P_{11}(k_3) F_2^{\rm (s)}(\vk_2,\vk_3)
 \int_{ \vq} P_{11}(q) F_3^{\rm (s)}(\vk_3,\vq,-\vq)
	+ \text{5 permutations}\ , 
	\label{eq:b321twoscoc} \\
B_{411}(k_1,k_2,k_3) &= 12P_{11}(k_2) P_{11}(k_3) \int_{\vq} P_{11}(q)
	F_4^{\rm (s)}(\vq,-\vq,-\vk_2,-\vk_3) + \text{2 permutations}
	\label{eq:b411scoc}
\end{align}
and we have used $\int_{\vq}\equiv \int \frac{d^3q}{(2\pi)^3}$.
However, as described in~\cite{Carrasco:2013sva}, explicitly for the case of loop corrections to the power spectrum, it is possible and indeed very convenient to group these terms together as a single integrand to avoid possible numerical issues in the evaluation of the integral. These issues could arise from large IR contributions in each diagram that, for IR-safe quantities, cancel in the final answer. To accomplish this, we first need to map each potential IR divergence~\footnote{%
These potential IR divergences will become true divergences in a no-scale universe ($P_{11}(k) \propto k^n$) with $n\geq -1$, and therefore the IR-safety procedure is absolutely necessary to obtain sensible results from the loop integrals in this case. In {\lcdm}, the linear power spectrum has a much steeper slope in the IR ($P_{11}(k) \sim k^1$ as $k\to 0$), which ensures IR convergence. However, this steepening happens only for $k\lesssim k_{\rm BAO}\simeq  0.01\hinvMpc$, which means that for the $k$'s of interest here, there will be large contributions peaked at $k\sim k_{\rm BAO}$ from each diagram that would cancel in the complete computation and therefore pose, at least potentially, a numerical challenge. The IR-safe integrand is automatically well behaved in the IR and so there is no large contribution from IR modes. We have checked in the two-loop power spectrum that the IR-safe integrand leads to a much stronger numerical efficiency~\cite{Carrasco:2013sva,Carrasco:2013mua}. We expect this to be the case also for the bispectrum, even though we have not compared the performance of the IR-safe integrand directly for this case.
} in momentum space to the origin ($\vq=0$). These potential divergences are located at points where the argument of a factor of $P_{11}$ can approach zero, keeping the external momenta fixed at some nonzero values.

In $B_{321}^{\rm (II)}$ and $B_{411}$, the only point where this can occur is $\vq\to 0$, so no re-mapping is necessary. In $B_{321}^{\rm (I)}$, if we examine the first permutation of external momenta that is explicitly written above, there are potential divergences at $\vq\to 0$ and $\vq\to\vk_2$; the second point can be mapped to the first one by the following manipulation, where $b_{321}^{\rm (I)}(\vq)$ (which also depends on the external momenta, but we suppress that dependence here) refers to the integrand of $B_{321}^{\rm (I)}$:
\begin{align}
\nn
\int_{\vq} b_{321}^{\rm (I)}(\vq) &= \int_{q<|\vk_2-\vq|} \frac{d^3q}{(2\pi)^3} b_{321}^{\rm (I)}(\vq) 
	+  \int_{q>|\vk_2-\vq|} \frac{d^3q}{(2\pi)^3} b_{321}^{\rm (I)}(\vq) \\ \nn
&= \int_{q<|\vk_2-\vq|} \frac{d^3q}{(2\pi)^3} b_{321}^{\rm (I)}(\vq) 
	+  \int_{\tilde{q}<|\vk_2-\tilde{\vq}|} \frac{d^3\tilde{q}}{(2\pi)^3} 
	b_{321}^{\rm (I)}(\vk_2-\tilde{\vq}) \\
&= 2 \int_{q<|\vk_2-\vq|} \frac{d^3q}{(2\pi)^3} b_{321}^{\rm (I)}(\vq)
	= 2 \int_{\vq} b_{321}^{\rm (I)}(\vq) \, \Theta(|\vk_2-\vq|-q)\ ,
\end{align}
where we have used the symmetry of $b_{321}^{\rm (I)}(\vq)$ under the substitution $\vq\to\vk_2-\vq$. An analogous manipulation must be applied to each of the other 5 permutations of external momenta.

The case of $B_{222}$ is more complicated, for two reasons. The first is the fact that divergences can occur at three locations ($\vq\to 0$, $\vq\to \vk_2$, and $\vq\to -\vk_3$), and the last two must be mapped to the first one. The second is that the form of $B_{222}$ written in Eq.~(\ref{eq:b222scoc}) is proportional to $P_{11}(k_2) P_{11}(k_3)$ in the $\vq\to 0$ limit of the integral, while there are permutations of the other terms that will be proportional to $P_{11}(k_1)$ in this limit. Since the IR limits of each integral must cancel with each other regardless of the form of $P_{11}$, this means that we might expect to have to manipulate $B_{222}$ to produce terms proportional to $P_{11}(k_1)$, in addition to re-mapping the potential divergences mentioned above.

Fortunately, these manipulations proceed in a fairly straightforward way. First, let us multiply the integrand of $B_{222}$ by a sum of products of step functions, like so:
\begin{align}
\nn
B_{222}(k_1,k_2,k_3) &= 8 \int_{\vq} P_{11}(q) P_{11}(|\vk_2-\vq|) P_{11}(|\vk_3+\vq|) \\ \nn
&\qquad \times F_2^{\rm (s)}(-\vq,\vk_3+\vq)
	F_2^{\rm (s)}(\vk_3+\vq,\vk_2-\vq) F_2^{\rm (s)}(\vk_2-\vq,\vq) \\ \nn
&\qquad \times \lb \Theta(|\vk_2-\vq|-q) 
	\left\{ \Theta(|\vk_3+\vq|-q) + \Theta(q-|\vk_3+\vq|) \right\} \right. \\
&\qquad\quad\quad \left. + \, \Theta(q-|\vk_2-\vq|) 
	\left\{ \Theta(|\vk_3+\vq|-q) + \Theta(q-|\vk_3+\vq|) \right\} \rb\ .
\label{eq:b222split}
\end{align}
We can now split the integrand into four terms, each multiplying a different product of step functions, and manipulate them so that each term has potential IR divergences only as $\vq\to 0$:
\begin{enumerate}
\item \underline{$\Theta(|\vk_2-\vq|-q) \Theta(|\vk_3+\vq|-q)$}:

As $\vq\to\vk_2$ or $\vq\to-\vk_3$, these step functions evaluate to zero, and therefore this term requires no further manipulation.
\item \underline{$\Theta(|\vk_2-\vq|-q) \Theta(q-|\vk_3+\vq|)$}:

This term is zero when $\vq\to\vk_2$, but not when $\vq\to-\vk_3$. However, if we let $\vq=-\vk_3-\tilde{\vq}$, then the integrand becomes
\begin{align}
\nn
&8P_{11}(|\vk_3+\tilde{\vq}|) P_{11}(|-\vk_1+\tilde{\vq}|) P_{11}(\tilde{q})
	F_2^{\rm (s)}(\vk_3+\tilde{\vq},-\tilde{\vq})
	F_2^{\rm (s)}(-\tilde{\vq},-\vk_1+\tilde{\vq}) F_2^{\rm (s)}(-\vk_1+\tilde{\vq},-\tilde{\vq}) \\
&\qquad \times \Theta(|-\vk_1+\tilde{\vq}|-|\vk_3+\tilde{\vq}|)
	\Theta(|\vk_3+\tilde{\vq}|-\tilde{q})\ ,
\end{align}
which has potential singularities at $\tilde{\vq}\to \vk_1$ and $\tilde{\vq}\to -\vk_3$, but the step functions evaluate to zero in these limits.
\item \underline{$\Theta(q-|\vk_2-\vq|) \Theta(|\vk_3+\vq|-q)$}:

This term is zero when $\vq\to-\vk_3$, but not when $\vq\to\vk_2$, but the substitution $\vq=\vk_2-\tilde{\vq}$ changes the integrand to
\begin{align}
\nn
&8P_{11}(|\vk_2-\tilde{\vq}|) P_{11}(\tilde{q}) P_{11}(|\vk_1+\tilde{\vq}|)
	F_2^{\rm (s)}(-\vk_2+\tilde{\vq},-\vk_1-\tilde{\vq})
	F_2^{\rm (s)}(-\vk_1-\tilde{\vq},\tilde{\vq}) F_2^{\rm (s)}(\tilde{\vq},\vk_2-\tilde{\vq}) \\
&\qquad \times \Theta(|\vk_2-\tilde{\vq}|-\tilde{q})
	\Theta(|\vk_1+\tilde{\vq}|-|\vk_2-\tilde{\vq}|)\ ,
\end{align}
which again only has potential singularities at $\tilde{\vq}\to 0$, thanks to the step functions.
\item \underline{$\Theta(q-|\vk_2-\vq|) \Theta(q-|\vk_3+\vq|)$}:

After substituting $\vq=\vk_2-\tilde{\vq}$, the integrand is the same in case 3, but with step functions
\beq
\Theta(|\vk_2-\tilde{\vq}|-\tilde{q}) \Theta(|\vk_2-\tilde{\vq}|-|\vk_1+\tilde{\vq}|)\ ,
\eeq
which evaluate to zero as $\tilde{\vq}\to \vk_2$, and also for $\tilde{\vq}\to -\vk_1$, but only if $k_3<k_1$. If $k_3>k_1$, we should repeat the steps above, starting from Eq.~(\ref{eq:b222split}), but interchanging $\vk_1$ and $\vk_3$ and propagating this change through each step. Each of the two cases can then be selected for via appropriate step functions (we will write this explicitly when we present the final $B_\text{1-loop}$ expression below).
\end{enumerate}
Thus, we have shown how to write the integrand of $B_{222}$ in a way that the only potential divergence is located at $\vq\to 0$. Notice also that our manipulations have introduced terms containing $P_{11}(k_1)$ in this limit, which will serve to cancel the corresponding terms appearing when external momenta are permuted in the other diagrams.

While the above procedure takes care of the leading potential IR divergences, there are still subleading divergences that would normally disappear after integration over orientations of $\vq$ is carried out. These can be made to cancel prior to integration if an appropriate symmetrization over angles is implemented, but the details of the symmetrization depend on the technique used to translate the integrand into a form that can be evaluated numerically~\footnote{%
For example, we could choose the $z$-axis to lie along $\vk_1$, and make the substitution $\vq=(q\sqrt{1-\mu^2}\cos(\phi),q\sqrt{1-\mu^2}\sin(\phi),q\mu)$. In this case, there are potentially IR-divergent terms containing $(\vk_2\cdot\vq)^2$, which itself contains the term $\sim \mu \sqrt{1-\mu^2} \sin(\phi)$. This dangerous term will vanish if the integrand is symmetrized in $\mu$ and $-\mu$, but {\em not} if the symmetrization is over $\vq$ and $-\vq$, since this would flip the signs of both $\mu$ and $\sin(\phi)$. In our own numerical work, we choose to symmetrize in $\mu$ and $-\mu$, and have verified analytically for the case of a scaling universe that all subleading IR divergences cancel as intended when this symmetrization is implemented.
}.

Thus, the final IR-safe form of $B_\text{1-loop}$ can be written in the following way:
\beq
B_\text{1-loop}(k_1,k_2,k_3) = \int_{\vq}
	\lb b_{222}^{(k_3>k_1)} \Theta(k_3-k_1)+b_{222}^{(k_3<k_1)} \Theta(k_1-k_3)
	+b_{321}^{\rm (I)}+b_{321}^{\rm (II)}+b_{411} \rb_\text{sym.\ over angles}\ ,
\eeq
where
\begin{align}
\nn
b_{222}^{(k_3>k_1)} &\equiv 8 P_{11}(q) P_{11}(|\vk_2-\vq|) P_{11}(|\vk_3+\vq|) F_2^{\rm (s)}(-\vq,\vk_3+\vq)
	F_2^{\rm (s)}(\vk_3+\vq,\vk_2-\vq) \\ \nn
&\qquad \times F_2^{\rm (s)}(\vk_2-\vq,\vq) \Theta(|\vk_2-\vq|-q) \Theta(|\vk_3+\vq|-q) \\ \nn
&\quad+8P_{11}(|\vk_3+\tilde{\vq}|) P_{11}(|-\vk_1+\tilde{\vq}|) P_{11}(\tilde{q})
	F_2^{\rm (s)}(\vk_3+\tilde{\vq},-\tilde{\vq})
	F_2^{\rm (s)}(-\tilde{\vq},-\vk_1+\tilde{\vq})  \\ \nn
&\qquad \times  F_2^{\rm (s)}(-\vk_1+\tilde{\vq},-\tilde{\vq})
	\Theta(|-\vk_1+\tilde{\vq}|-|\vk_3+\tilde{\vq}|) \Theta(|\vk_3+\tilde{\vq}|-\tilde{q}) \\ \nn
&\quad +8P_{11}(|\vk_2-\tilde{\vq}|) P_{11}(\tilde{q}) P_{11}(|\vk_1+\tilde{\vq}|)
	F_2^{\rm (s)}(-\vk_2+\tilde{\vq},-\vk_1-\tilde{\vq})
	F_2^{\rm (s)}(-\vk_1-\tilde{\vq},\tilde{\vq}) \\ \nn
&\qquad \times F_2^{\rm (s)}(\tilde{\vq},\vk_2-\tilde{\vq})  \Theta(|\vk_2-\tilde{\vq}|-\tilde{q})
	\Theta(|\vk_1+\tilde{\vq}|-|\vk_2-\tilde{\vq}|) \\ \nn
&\quad +8P_{11}(|\vk_2-\tilde{\vq}|) P_{11}(\tilde{q}) P_{11}(|\vk_1+\tilde{\vq}|)
	F_2^{\rm (s)}(-\vk_2+\tilde{\vq},-\vk_1-\tilde{\vq})
	F_2^{\rm (s)}(-\vk_1-\tilde{\vq},\tilde{\vq}) \\
&\qquad \times F_2^{\rm (s)}(\tilde{\vq},\vk_2-\tilde{\vq})  \Theta(|\vk_2-\tilde{\vq}|-\tilde{q})
	\Theta(|\vk_2-\tilde{\vq}|-|\vk_1+\tilde{\vq}|)\ , \\
b_{222}^{(k_3<k_1)} &\equiv \left. b_{222}^{(k_3>k_1)}\right|_{\vk_1\leftrightarrow\vk_3}\ ,
\end{align}
and
\begin{align}
\nn
b_{321}^{\rm (I)} &\equiv 12P_{11}(k_3)  P_{11}(q) P_{11}(|\vk_2-\vq|) 
	F_3^{\rm (s)}(-\vq,-\vk_2+\vq,-\vk_3) F_2^{\rm (s)}(\vq,\vk_2-\vq) \\
	&\qquad \times \Theta(|\vk_2-\vq|-q) + \text{5 permutations}   \ , \\
b_{321}^{\rm (II)} &\equiv 6P_{11}(k_2) P_{11}(k_3) F_2^{\rm (s)}(\vk_2,\vk_3)
	 P_{11}(q) F_3^{\rm (s)}(\vk_3,\vq,-\vq) + \text{5 permutations}\ , \\
b_{411} &\equiv 12P_{11}(k_2) P_{11}(k_3) P_{11}(q)
	F_4^{\rm (s)}(\vq,-\vq,-\vk_2,-\vk_3) + \text{2 permutations}\ .
\end{align}

\subsection{Counterterms}
\label{sec:counterterms}

In the EFTofLSS, the effects of nonlinearities at short distance scales are encapsulated into an effective stress tensor, $\gammai{}^i$, on the right-hand side of the Euler equation.\footnote{%
The operator $\gammai{}^i$ is defined by $\gammai{}^i\equiv \rho^{-1} \d_j \tau^{ij}$, which is the quantity that actually appears in the Euler equation, where $\tau^{ij}$ is the form of the stress tensor more commonly seen in standard treatments of fluids. It is somewhat of an arbitrary choice whether to write an expansion for $\tau^{ij}$ or $\rho^{-1} \d_j \tau^{ij}$; we find it more convenient to expand the latter, and for brevity we will sometimes refer to $\gammai{}^i$ as the ``stress tensor," although technically this is a mild abuse of terminology.
} This stress tensor is written as a double expansion in fields and derivatives, consistent with the symmetries that are relevant at the appropriate scales (namely, rotational invariance and the equivalence principle) The various terms in this expansion add extra contributions to the perturbative solutions for $\delta$ and $v_i$, which lead to the appearance of counterterms in expressions for various observables.

As noted in~\cite{Carrasco:2013mua,Carroll:2013oxa}, there is no obvious separation of timescales between the evolution of short and long modes of the density and velocity fields, and therefore $\gammai{}^i$ may exhibit significant non-locality in time. This non-locality will manifest itself in counterterms that are evaluated at loop level, such as those appearing in a two-loop calculation of the power spectrum. In the tree-level counterterms that we restrict ourselves to in this work, any non-locality in time will only have a very minor effect. Furthermore, since in~\cite{Carrasco:2013mua} we found that nonlinear power spectrum data seem to prefer only a mild degree of non-locality at best, we will comment on where this non-locality might play a role, but in concrete calculations we will take the local-in-time limit for simplicity.

The lowest-order term in the stress tensor is the linear one, which in the local-in-time limit looks like a ``speed of sound" term: schematically,
\beq
\label{eq:speedofsoundterm}
\gammai{}^i \supset c_{\rm s}^2 \d^i \d^2\phi \sim c_{\rm s}^2 \d^i \delta\ .
\eeq
At tree-level, this term will contribute two counterterms to the bispectrum (see App.~\ref{sec:eftreview} for details of the derivation), which, when grouped together into a single expression, take the form
\begin{align}
\nn
B_{c_{\rm s}}(k_1,k_2,k_3) &= 2 P_{11}(k_1) P_{11}(k_2) \tilde{F}_2^{\rm (s)}(\vk_1,\vk_2)
	+ \text{2 permutations} \\
&\quad - 2 \bar{c}_1 k_1^2 P_{11}(k_1) P_{11}(k_2) F_2^{\rm (s)}(\vk_1,\vk_2)
	+ \text{5 permutations}\ ,
\label{eq:bcsdef}
\end{align}
where $\tilde{F}_2^{\rm (s)}(\vk_1,\vk_2)$ is given by Eq.~(\ref{eq:f2stilde}) in App.~\ref{sec:eftreview}.  $\bar{c}_1$ is related to the free parameter $\co$ (which also enters into the one-loop EFT prediction for the power spectrum) by $\bar{c}_1 = (2\pi) \co/\knl^2$, while $\tilde{F}_2^{\rm (s)}$ is proportional to a different constant $\bar{c}_2$, which itself is equal to $\bar{c}_1$ times some order-one factor related to the severity of the non-locality in time. Unless the non-locality is very severe, this factor will be very close to one, so the assumption of locality in time (for which $\bar{c}_2=\bar{c}_1$ exactly) will not have a huge effect.

There are also four other possible tree-level counterterms we can write down, arising from the contribution to the effective stress tensor that is quadratic in the fields. Three of the terms arise from operators like $(\d^2\phi)^2$: schematically, we have 
\begin{align}
\nn
\gammai{}^i &\supset
\d^i \lb \d^2\phi \rb^2 
	+  \d^i  \lb \d^j \d^k\phi \, \d_j \d_k \phi \rb
	+  \d^i \d^j\phi\, \d_j \d^2 \phi  \\
&\sim \d^i \delta^2 + 
	\d^i \lb \frac{ \d^j \d^k}{\d^2} \delta \cdot\frac{ \d_j \d_k }{\d^2}\delta \rb +
	 \frac{ \d^i\d^j}{\d^2} \delta \cdot \d_j \delta\ ,
\label{eq:stress3cts}
\end{align}
where we used Poisson's equation to trade $\d^2\phi$ for $\delta$ in the second line. Recall that we could also have written a linear counterterm involving the velocity divergence:
\beq
\gammai{}^i \supset \d^i (\d_j v^j)\ ,
\eeq
At linear order, we have that $\theta=\d_i v^i=-\H f \delta$, where $f \equiv \partial \log D_1/\partial \log a$ with $\H=a H$ and $D_1$ being the linear growth factor. As usual in SPT, we can rewrite the velocity field as (see Eq.~(\ref{eq:thetaexp}) for a more detailed discussion) 
\bea
\label{eq:thetaexptemp}
\theta(a,\vk) &=& -\H(a)\, f\, \hat{\theta}\ , \quad{\rm where}\quad \hat{\theta}(a,\vk)= \sum_{n=1}^\infty \, [D_1(a)]^n \theta^{(n)}(\vk) \ .
\eea
So we define the new counterterm as proportional to 
\beq
\gammai{}^i \supset \d^i (\hat{\theta} - \delta)\ ,
\eeq
which starts at second order in the fluctuations.

These four counterterms take respectively the following form:
\begin{align}
\label{eq:bc1}
B_{c_1}(k_1,k_2,k_3) &= -2c_1 \, k_1^2 P_{11}(k_2) P_{11}(k_3) + \text{2 permutations}\ , \\ \nonumber
B_{c_2}(k_1,k_2,k_3) &= -2c_2 \, k_1^2 \frac{(\vk_2\cdot\vk_3)^2}{k_2^2 k_3^2} 
	P_{11}(k_2) P_{11}(k_3) + \text{2 permutations}\ , \\ \nonumber
B_{c_3}(k_1,k_2,k_3) &= -c_3 \, (\vk_2\cdot\vk_3) \lb \frac{\vk_1\cdot\vk_2}{k_2^2} 
	+ \frac{\vk_1\cdot\vk_3}{k_3^2} \rb P_{11}(k_2) P_{11}(k_3) + \text{2 permutations}\ , \\ \nonumber
B_{c_4}(k_1,k_2,k_3)& = -c_4  \,k_1^2 \, \frac{k_1^4 + (k_2^2-k_3^2)^2 - 2k_1^2 (k_2^2+k_3^2)}{7k_2^2 \, k_3^2} 
	P_{11}(k_2) P_{11}(k_3) + \text{2 permutations}\ ,
\end{align}
where $c_1,\dots,c_4$ are arbitrary coefficients, each expected to be of order $\knl^{-2}$ (there are no factors of~$(2\pi)$ because these counterterms are associated with loop diagrams that do not contain any trivial angular integrations). Notice that $B_4$ is a linear combination of $B_1$ and $B_2$
\be
B_{c_4}(k_1,k_2,k_3)=-\frac{2}{7}c_4 \left(\frac{B_1(k_1,k_2,k_3)}{c_1}-\frac{B_2(k_1,k_2,k_3)}{c_2}\right)\ ,
\ee
therefore, since $c_1$ and $c_2$ are free parameters, we can neglect $B_{c_4}$ from the rest of the paper.

\begin{figure}[t]
\begin{center}
\includegraphics[width=0.45\textwidth]{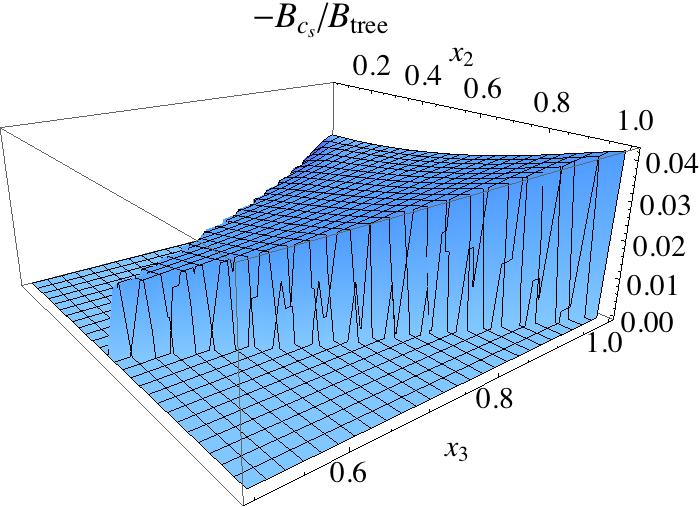}
\includegraphics[width=0.45\textwidth]{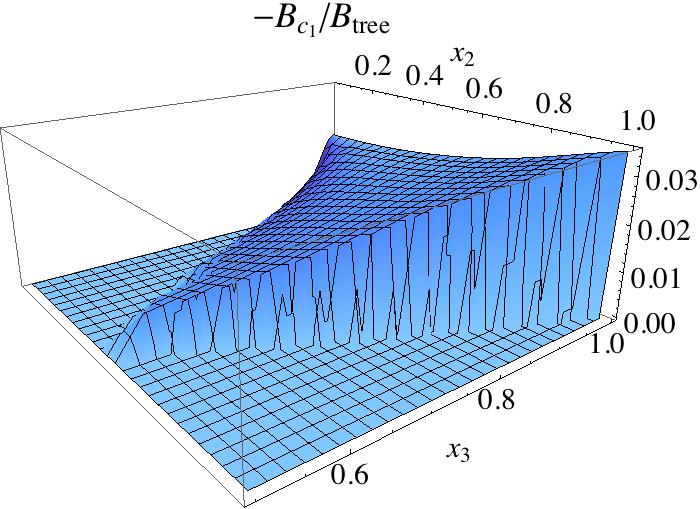}
\includegraphics[width=0.45\textwidth]{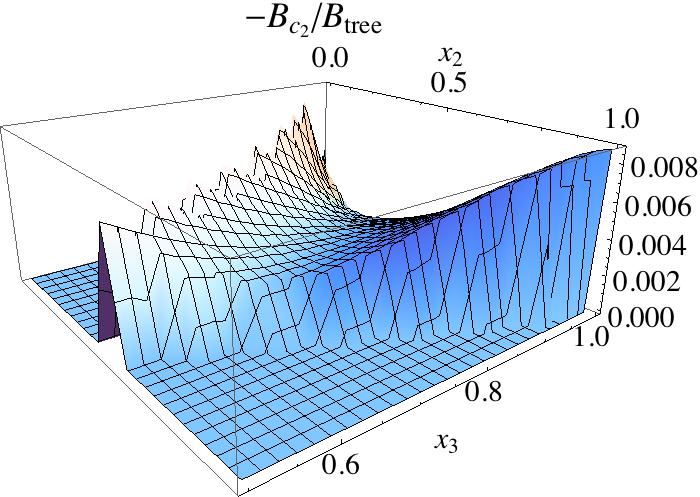}
\includegraphics[width=0.45\textwidth]{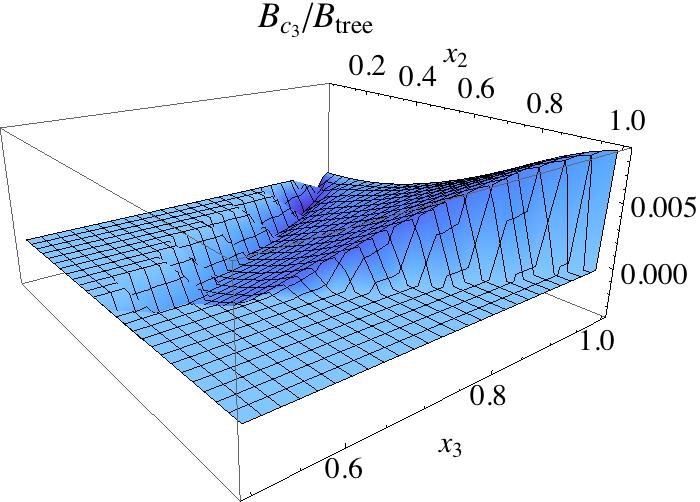}
\caption{ \label{fig:shape-plots} \footnotesize\it  Ratios of various terms in the bispectrum prediction to the tree-level expression $B_\text{tree}$, plotted with $k_1=0.1\hinvMpc$ and in terms of $x_2\equiv k_2/k_1$ and $x_3\equiv k_3/k_1$. To avoid redundancy, we only plot configurations with $x_2\leq x_3$, while the triangle inequality restricts physical configurations to satisfy $1-x_3\leq x_2$. Each term is strongest on equilateral triangles ($x_2=x_3=1$), becoming relatively weaker for other configurations such as squeezed ($x_2\to 0$) or flat ($x_2+x_3=1$). This implies that configurations where three short modes interact are more nonlinear than configurations involving one or more longer modes and one short mode---in some sense, of all triangles with $k_1$ fixed, equilateral triangles are ``closest" to the nonlinear scale.  As we let $k_1$ grow, all terms grow in size relative to $B_{\rm tree}$ but  the shape remains quite unaltered. }
\end{center}
\end{figure}

\begin{figure}[ht]
\begin{center}
\includegraphics[width=0.45\textwidth]{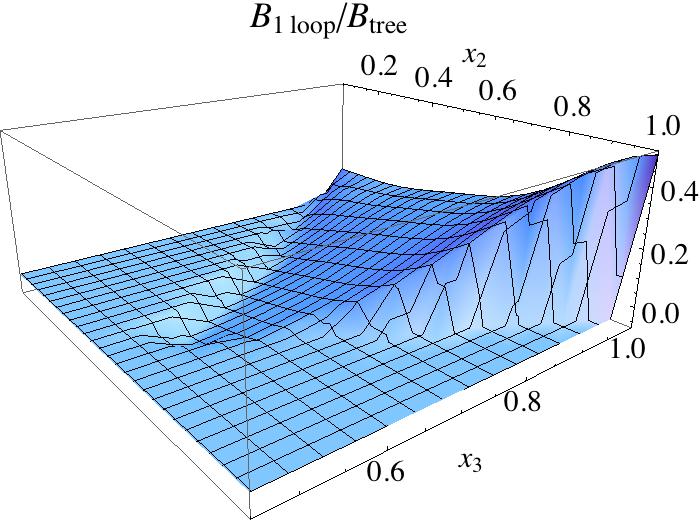}
\includegraphics[width=0.41\textwidth]{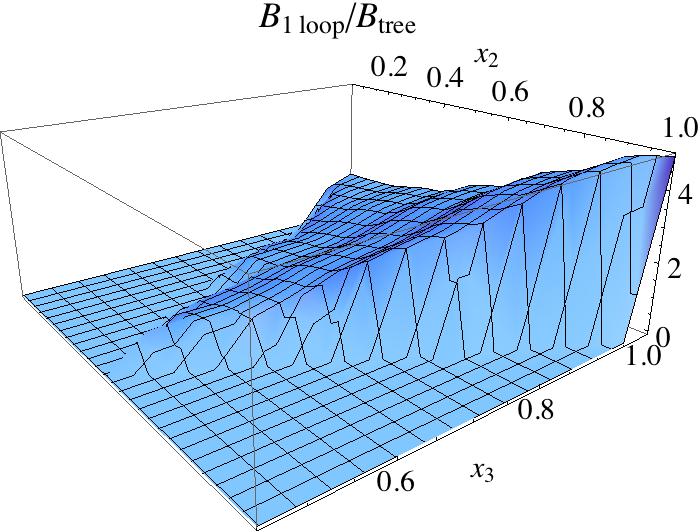}
\caption{ \label{fig:shape-plots-2} \footnotesize\it  Same as Fig.~\ref{fig:shape-plots}, but for the 1-loop contribution for $k_1=0.1 \hinvMpc$ on the left and  $k_1=0.3 \hinvMpc$ on the right.}
\end{center}
\end{figure}

To gain insight into the effect of these terms on the bispectrum in various configurations, in Fig.~\ref{fig:shape-plots} we plot the ratio of each  counterterm to $B_\text{tree}$ as a function of $x_2\equiv k_2/k_1$ and $x_3\equiv k_3/k_1$ for $k_1=0.1\hinvMpc$ (the scale-dependence of each term is such that these plots do not change significantly for other values of $k_1$).  In Fig.~\ref{fig:shape-plots-2}, we show $B_\text{1-loop}/B_\text{tree}$ for $k_1=0.1\hinvMpc$ and $0.3\hinvMpc$. To restrict to physical triangles and avoid redundancy in the plots, we only plot in the range $1-x_3<x_2<x_3$. We see from these plots that each term is strongest for equilateral triangles ($x_2=x_3=1$), and they each become less important for squeezed ($x_3\approx 1$, $x_2\approx 0$) and flat ($x_2+x_3\approx 1$) configurations (except for $B_{c_2}$, which tends to peak somewhat in the flat limit). This matches our intuition that of all triangles with one side fixed, equilateral triangles are the most nonlinear, because they are the ones for which each side is closest to the nonlinear scale.

In App.~\ref{sec:renormalization} we show that in the case in which the universe is described by a simple power law, these three counterterms are required to ensure that UV divergences can be cancelled in the bispectrum. This is a sufficient but non-necessary condition that tells that these counterterms should be present in the EFT. It is not a necessary condition because even if these terms were not needed to cancel any UV divergence, they would still be allowed by the symmetry of the problem and therefore should be included.

\section{Comparison with Simulations}
\label{sec:comparison}

\subsection{N-body simulations}
\label{sec:simulations}

We compare the predictions of the EFTofLSS against the results of a
dark-matter only $N$-body simulation. We use the Millennium-XXL (MXXL) 
simulation \cite{Angulo2012}, which evolved $6720^3$ particles inside a 
comoving cubical region of $L_{\rm box} = 3000\,h^{-1}\,$Mpc a side, 
from $z=63$ to the present day. The combination of a large volume with 
a high mass resolution of the MXXL suppresses the effects of cosmic 
variance and discreteness noise in the bispectrum estimates, therefore
allowing a detailed comparison with EFT predictions.

The MXXL simulation was carried out with the following set of cosmological
parameters: a matter density of $\Omega_\mathrm{m}=0.25$ in units of the 
critical density, a cosmological constant with $\Omega_\Lambda=0.75$, a Hubble
constant $h=0.73$ in units of $100\, \mathrm{km}\,\mathrm{s}^{-1}\Mpc^{-1}$, a
spectral index $n_s=1$ and a normalization parameter $\sigma_8=0.9$ for the
primordial linear density power spectrum. The initial linear theory power 
spectrum was computed using the Boltzmann code CAMB~\cite{Lewis:1999bs}, 
 and the initial particle positions were determined  by adding displacements 
given by second-order Lagrangian perturbation theory to a random glass-like 
configuration.
Exactly the same power spectrum and cosmological parameters were used in 
our EFT calculations.

\subsubsection{Bispectrum and estimation of errors}

In this paper we focus on the bispectrum of the nonlinear mass density 
contrast at $z=0$. In order to compute this quantity we first construct
a density field by assigning simulation particles onto a cubic grid
using a ``cloud-in-cells" deposit scheme. We employ a grid of $2048^3$ 
points 
and then Fast Fourier Transform the respective density field. Finally,
we correct for the effects of the assignment scheme by dividing each
Fourier mode by the Fourier transform of a cubical top-hat window.

The power spectrum is computed by spherically averaging the amplitude
of Fourier modes in annuli of radius $\delta k$.
The bispectrum, $B(k_1, k_2, \theta)$, is obtained by performing a nested
loop over all grid points ($2048^6$) 
and averaging $\delta(k_1) \delta(k_2) \delta(k_3)$ over triangles whose 
sides satisfy
the following condition: $k_1^i = k_1 \pm \frac{1}{2} \delta k$, 
$k_2^i = k_2 \pm \frac{1}{2} \delta k$ and $\theta^i = \theta \pm \frac{1}{2} \delta \theta$,
where $\theta^i = \cos^{-1}\left( \hat{k}_1^i \cdot \hat{k}_2^i \right)$.
We note that in subsequent comparisons we discard the imaginary part of the
measured bispectrum, and consider only the real one.
For these calculations we set $\delta k = 2\pi/L_{\rm box} = 0.0021\,h^{-1}$Mpc, 
and $\delta \theta = \pi/20$.

Specifically, we use triangular configurations with $k_1$ and $k_2$ given by the following list:
\begin{align}
k_2 = k_1:& \qquad k_1 \in \{ 0.04, 0.06, \dots, 0.38 \} \invMpc\ , \\
k_2 = 1.5k_1:& \qquad  k_1 \in \{ 0.06, 0.08, \dots, 0.22 \} \invMpc\ , \\
k_2 = 2k_1:& \qquad k_1 \in \{ 0.06, 0.10, 0.14, 0.18, 0.22 \} \invMpc\ .
\end{align}
For each $(k_1,k_2)$ pair above, we consider 19 different triangles, determined by the angle between $k_1$ and $k_2$, for which we use 19 equally-spaced values between $0$ and $\pi$. This gives 608 total bispectrum data points, although our method of fitting to triangles with maximum side length less than some cutoff $k_{\rm max}$ (see Sec.~\ref{sec:results}) means that we only use certain subsets of these data points in our final analysis. In particular, there are 289 triangles with maximum side length less than $0.27\invMpc$.

To close this section we discuss the estimation of the error in the 
measurement of the bispectrum. We adopt the commonly-used expression
derived by \cite{Fry1993,Feldman1994,Scoccimarro:1997st,Scoccimarro2004}:

\beq
\label{eq:biserrcv}
[\Delta B(k_1,k_2,\theta)]^2 = L_{\rm box}^3 \frac{s_{123}}{n_{\rm triangles}} P_{\rm NL}(k_1)P_{\rm NL}(k_2)P_{\rm NL}(k_3)\ .
\eeq

\noindent This formula is derived in the Gaussian limit, where $P_{\rm NL}$ should be replaced by $P_{11}$, but it seems to be a better approximation for
the  scales we are interested in here to replace it with $P_{\rm NL}$. In this expression: $s_{123}$
is equal to 6, 2, 1 for equilateral, isosceles and general triangles, respectively; 
$n_{\rm triangles}$ is the number of triangles contributing to a given 
configuration, which, in our case, is directly counted inside the bispectrum
code; and $k_3^2 = k_1^2 + k_2^2 - 2 k_1 k_2 \cos(\pi-\theta)$
.  On top of this error, we add an extra 2\% error for each bispectrum data point, to account for possible unknown systematic errors in the simulation and in the comparison between simulations and the EFT.

\subsection{Determining $\co$ from the matter power spectrum}
\label{sec:csfit}

The one-loop EFT prediction for the bispectrum will involve the parameter $\co$, which can be determined by fitting the one-loop power spectrum prediction to nonlinear data. As discussed in~\cite{Carrasco:2012cv,Carrasco:2013mua,Senatore:2014via}, the one-loop power spectrum in the EFTofLSS takes the form
\beq
\label{eq:peft1loop}
P_{\text{EFT-1-loop}} = P_{11} + P_{\text{1-loop}} -  { 2\, (2\pi)} \co \frac{k^2}{\knl^2} P_{11} \ ,
\eeq
and is expected to be a valid match to the data up to the wavenumber when the two-loop contribution becomes non-negligible. By fitting Eq.~(\ref{eq:peft1loop}) to the power spectrum measured from the simulations described in Sec.~\ref{sec:simulations}, over the range $0.02\invMpc \leq k \leq 0.1 \invMpc$, we find
\beq  
\label{eq:cofromps}
\co =( 1.52 \pm {  0.56}) \times \frac{1}{2\pi} \lp \frac{\knl}{\invMpc} \rp^{2} \qquad (\text{1-$\sigma$}) .  
\eeq
The upper limit of the fit range is chosen to be roughly where the one-loop prediction begins to deviate significantly from the nonlinear data~\footnote{%
Other choices of the fit range are possible, but the results for the performance of the bispectrum prediction are not significantly affected by this choice. For instance, if $\co$ is fit over the range $0.15\invMpc<k<0.25\invMpc$, as in~\cite{Carrasco:2013mua}, we find $\co \approx 2.39/(2\pi) (\knl/[\invMpc])^2$, but using this value of $\co$ in the bispectrum prediction only changes the reach of the prediction by $\sim$10\%, from $k\sim 0.25\invMpc$ to $k\sim 0.27\invMpc$. (See Sec.~\ref{sec:results} for a discussion of how this reach is defined and determined.) Such a change in $\co$ will of course affect the performance of the power spectrum prediction, but we leave an investigation of this point to future work.
}.

\begin{figure}[t]
\begin{center}
\includegraphics[width=0.8\textwidth]{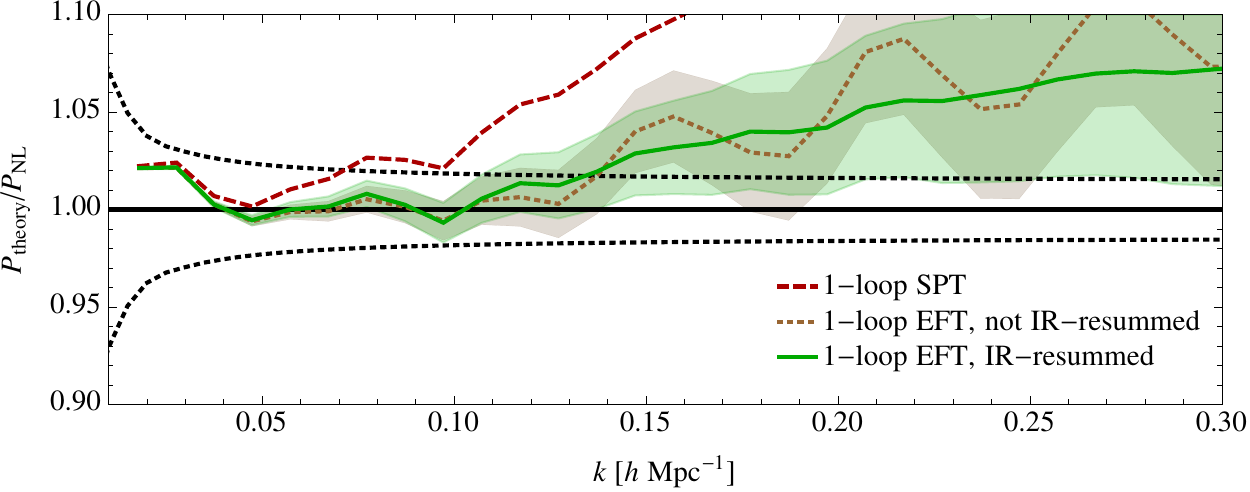}
\caption{ \label{fig:psfit}  \footnotesize\it  One-loop SPT and EFT predictions for the matter power spectrum, normalized to measurements from the simulations described in Sec.~\ref{sec:simulations}. The parameter $\co$ in the EFT prediction has been determined by fitting the IR-resummed curve over the range $0.02\hinvMpc \leq k \leq 0.1\hinvMpc$.
The resulting value for $\co$ can then be used in the EFT prediction for the bispectrum, without re-fitting. The  black dashed line corresponds to cosmic variance plus an assumed 1\% systematic error in the simulations, added in quadrature with a 1\% accuracy goal for the prediction. The shaded bands show the uncertainty in the EFT predictions from the 1$\sigma$ uncertainty in the value of $\co$.}
\end{center}
\end{figure}

The results of the fit are shown in Fig.~\ref{fig:psfit}. In the fit, we use the sampling variance error on $P_{\rm NL}$, and also add an extra 1\% errorbar to account for possible unknown systematic errors in the simulation and in the comparison between simulations and the EFT. As in~\cite{Carrasco:2013mua}, there are residual oscillations of $\sim$2\% around the mean of the prediction. These oscillations are due to the effects of bulk flows, and can be eliminated by an appropriate resummation of IR modes, as described in detail in~\cite{Senatore:2014via}~\footnote{%
We find a way to slightly simplify the numerical implementation of the techniques in~\cite{Senatore:2014via}. We describe it in App.~\ref{app:IRresum}.
} and shown in green in the same figure~\footnote{%
There seems to be a slight systematic offset between the IR-resummed and non-resummed EFT predictions. This is a very small, quite inconsequential, effect, of about 1\% in size, whose origin is hard to investigate. We leave this to future work.
}.

\subsection{Bispectrum estimates}
\label{sec:estimates}

In order to utilize the EFTofLSS to make a prediction about the matter bispectrum (or, indeed, any observable), one should have some way of determining which loop corrections and counterterms should be included to reach a given accuracy goal. In~\cite{Carrasco:2013mua}, estimates were made about the sizes of various terms by treating them in the scaling universe approximation, with an effective $\knl$ and the tilt of the linear power spectrum fit over the range in which the prediction was being made.

We can perform a similar exercise for the bispectrum, in order to estimate when the two-loop correction might become important. The most efficient way to do so is to determine which diagrams will have the largest contribution, and then make a more detailed estimate of their sizes. These diagrams will arise from the correlator $\left\la \delta^{(3)} \delta^{(3)} \delta^{(2)} \right\ra$, and will scale in the following way when evaluated on equilateral configurations:
\beq
\label{eq:b332}
B_{332} \sim \frac{192}{3!\, 3!\, 2!} (2\pi)^2 \lp \frac{k}{\knl} \rp^{2(3+n)} P_{11}(k)^2\ .
\eeq
The $(2\pi)^2$ arises from the fact that these diagrams contains two loop integrals that are not nested, which each contribute a factor of $(2\pi)$ (in the language of~\cite{Carrasco:2013mua}, they are ``reducible" diagrams). The~192 is a combinatoric factor related to the number of ways to sew together the diagrams' internal lines, counted by using diagrams with cubic vertices only (e.g.~\cite{Crocce:2005xy,exacttimedep}). The factorials in the denominator arise from integrals over Green's functions in time: one can consider the Einstein de-Sitter case, when these Green's functions are roughly the scale factor~$a$, and each factor of $\delta^{(m)}$ will involve $m$ integrals over these factors, yielding a final prefactor of $\sim 1/m!$. The power of $(k/\knl)^{(3+n)}$ counts the number of loops, where $n$ is the approximate tilt of $P_{11}$ in the region we're making predictions for.

Other two-loop diagrams can be estimated in the same fashion, but they will all either have fewer factors of $(2\pi)$ or else will be more strongly suppressed by inverse factorial factors (for which the combinatorics cannot compensate). We should compare this estimate to $B_\text{EFT-1-loop}$, which we write as $B_\text{tree}+B_\text{1-loop}+B_{c_s}$. We approximate $B_\text{1-loop}$ by the size of its largest diagrams (in the sense of having the most factors of $(2\pi)$ along with the smallest amount of combinatorial suppression), which come from~$\left\la \delta^{(4)} \delta^{(1)} \delta^{(1)} \right\ra$ and $\left\la \delta^{(3)} \delta^{(2)} \delta^{(1)} \right\ra$:
\beq
\label{eq:b1loopest}
B_\text{1-loop} \sim \lp \frac{48}{4!} + \frac{72}{3!\, 2!}\rp (2\pi) \lp \frac{k}{\knl} \rp^{3+n} P_{11}(k)^2\ .
\eeq
For $B_{c_s}$, we count all the permutations in Eq.~(\ref{eq:bcsdef}):
\beq
\label{eq:bcsest}
B_{c_s} \sim 9\times  \frac{2}{2!} (2\pi) \co \frac{k^2}{\knl^2} P_{11}(k)^2\ .
\eeq
Finally, for $B_\text{tree}$, we use
\beq
\label{eq:btreeest}
B_\text{tree} \sim \frac{6}{2!} P_{11}(k)^2\ .
\eeq

To apply these estimates to the universe simulated in Sec.~\ref{sec:simulations}, we fit a scaling universe power spectrum to the linear spectrum corresponding to the simulations, and find that over the range $0.25\hinvMpc \lesssim k \lesssim 0.60 \hinvMpc$, $P_{11}(k) \approx \frac{(2\pi)^3}{\knl^3} \lp \frac{k}{\knl} \rp^n$ with $n\approx-2.1$ and $\knl\approx 3.2 \hinvMpc $. Therefore, at $k\sim 0.3\hinvMpc$,
\beq
\frac{B_\text{2-loop}}{B_\text{tree}+B_\text{1-loop}+B_{c_s}} \sim 0.15\ .
\eeq
This estimate is  uncertain up to $\mathcal{O}(1)$ factors, but it nevertheless indicates that $B_\text{2-loop}$ may strongly limit how far into the UV our prediction can reach.  Note that the same estimate applied to the universe from~\cite{Carrasco:2013mua}, which has a more realistic normalization of the power spectrum and hence a more realistic value of $\knl$, gives $0.09$ instead of $0.15$, implying that in the real universe, the reach of the prediction should be farther than what it is in this paper.

As discussed in Sec.~\ref{sec:counterterms}, there are also three tree-level counterterms besides $B_{c_s}$ that could conceivably be important for the one-loop bispectrum prediction. A scaling-universe estimate for the size of each of these terms yields $B_{c_i}/B_\text{2-loop}\sim0.09$ at $k\sim 0.3\hinvMpc$~\footnote{%
This estimate arises from the following steps. From Eq.~(\ref{eq:bc1}), we use 
\beq
\label{eq:bc1est}
B_{c_1} \sim 3\times 2 c_1 \, k^2 P_{11}(k)^2\ ,
\eeq
and comparing this with the estimate for $B_{332}$ from Eq.~(\ref{eq:b332}), we find
\beq\label{eq:c1vs2loop}
\frac{B_{c_1}}{B_\text{2-loop}} \sim \frac{9}{16\pi^2} \frac{c_1 \, k^2}{(k/\knl)^{2(3+n)}}\ .
\eeq
We expect $c_1 \sim \bar{c}_1/(2\pi)$, and upon plugging this in, along with $\bar{c}_1=1.52 {\rm Mpc}^2/h^2$, $\knl=3.2\hinvMpc$, and $n=-2.1$, we find a value of $\sim$0.09 at $k\sim 0.3\hinvMpc$. We expect similar contributions from $B_{c_2}$ and $B_{c_3}$.
}, implying that the additional counterterms are not parametrically larger than $B_\text{2-loop}$ in this region. Since, by definition, our one-loop prediction neglects $B_\text{2-loop}$, this argument implies that we are justified in neglecting these three counterterms as well~\footnote{%
Notice the following important fact. The current universe can be approximated with the union of two power laws, the transition point being at $\ktr\simeq 0.25\hinvMpc$~\cite{Carrasco:2013mua}. All our estimates so far have been focussed on the most UV region of our fit, where $n=-2.1$ and $\knl\simeq 3.2\hinvMpc$. For $k\lesssim \ktr$, the linear power spectrum is fitted with a power law with $n\simeq -1.7$ and $\tknl\sim 1.5\hinvMpc$. Eq.~(\ref{eq:c1vs2loop}) can be applied to $k$'s below $\ktr$ by using $\tknl$ and  $n\simeq -1.7$, a slope which is quite steeper than $-2.1$. This means that, as we move towards the infrared, the importance of the additional counterterms relative to $B_\text{2-loop}$ grows, and one could potentially find oneself in the situation in which one is allowed to include the additional counterterms without including  $B_\text{2-loop}$, because this is justified in some relatively IR region, while not being so in the most UV region. One can easily estimate that we are currently not in this situation for two concurrent reasons: first, by $k\simeq0.1\hinvMpc$, the ratio from (\ref{eq:c1vs2loop}) is still sensibly smaller than one;  second, as we move to the IR, the numerical data have larger error bars that make the inclusion of the additional counterterms, as well as of the two-loop term, practically unnoticeable, even though their effect might be at percent level. This interpretation is verified by Figures~\ref{fig:fits-isos},~\ref{fig:fits-3o2} and~\ref{fig:fits-2} below. This subtlety will become more apparent after we perform the two-loop computation, something that we plan to do in the future, and when we will have more precise data. In particular, with precise enough data, we expect the fit to improve also at low $k$'s below $\ktr$, after the inclusion of $B_\text{2-loop}$ and of the additional counterterms, a fact that with our available data is not measurable.
}. In the next section we will check this argument by including these counterterms in a fit to simulation data and examining their affect on the performance of the prediction: if they are smaller than $B_\text{2-loop}$, they should not help the fit too much.

\subsection{Results}
\label{sec:results}

In this section, we compare various bispectrum predictions to the measurements described in Sec.~\ref{sec:simulations}. To assess the goodness of fit, we calculate a $\chi^2$ statistic,
\beq
\chi^2(k_{\rm comp}) = \sum_{\Delta_i} 
	\frac{\left(B_{\rm data}(\Delta_i)-B_\text{theory}(\Delta_i)^2\right)}{\sigma_i^2}\ ,
\eeq
where $\Delta_i$ denotes each triangle of wavevectors with maximum side length less than $k_{\rm comp}$  and $\sigma_i$ is taken from Eq.~(\ref{eq:biserrcv}), and then integrate a $\chi^2$ distribution with the proper number of degrees of freedom from $\chi^2(k_{\rm comp})$ to infinity to find an effective $p$-value, indicating our confidence that the data are described well by each prediction.

\begin{figure}[t]
\begin{center}
\includegraphics[width=0.89\textwidth]{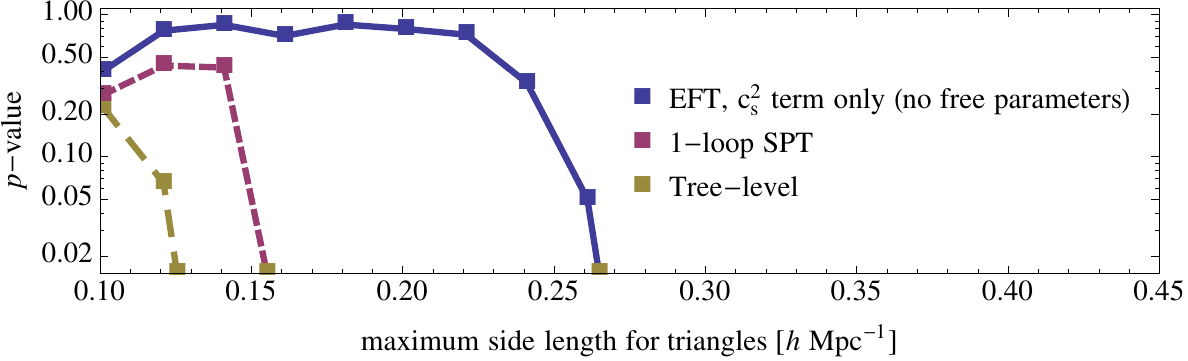}\\[0.3cm]
\includegraphics[width=0.89\textwidth]{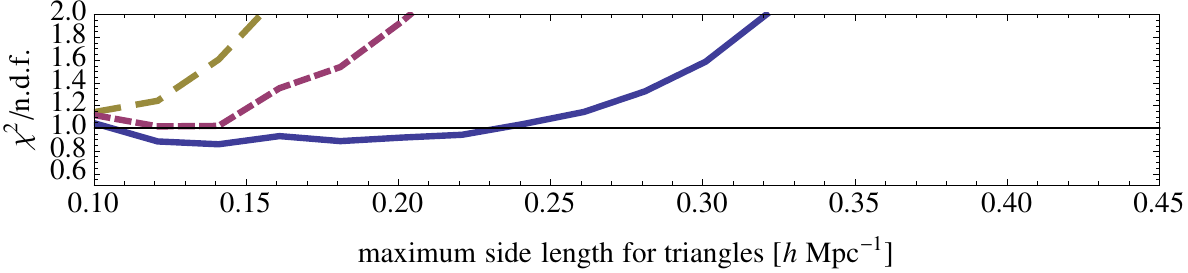}\\[0.3cm]
\includegraphics[width=0.89\textwidth]{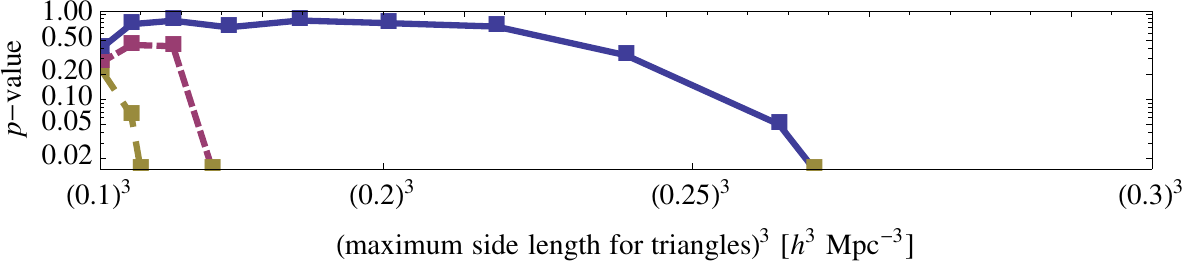}
\caption{ \label{fig:pval-kmax}  \footnotesize\it 
Top: $P$-values corresponding to comparisons of various theory curves to nonlinear data, as described in the main text, as a function of the maximum side length  ($k_{\rm comp}$) of the triangles used to compute the $p$-value. The solid, short-dashed, and long-dashed lines correspond to one-loop EFT, one-loop SPT, and tree-level SPT or EFT respectively. For the one-loop EFT prediction, the value of $\co$  has been fixed by fitting the EFT prediction for the matter power spectrum, so there is no free parameter. Middle: reduced $\chi^2$ for the various predictions. (It has the same information of the top panel, but presented in a different way.) Bottom: same as top, but with $k^3_{\rm comp}$ on the $x$-axis. Since the number of available modes grows as  $k^3_{\rm comp}$, the bottom plot gives a pictorial representation of the gain in information that is obtained by reaching higher $k$'s. }
\end{center}
\end{figure}

\begin{figure}[t]
\begin{center}
\includegraphics[width=0.45\textwidth]{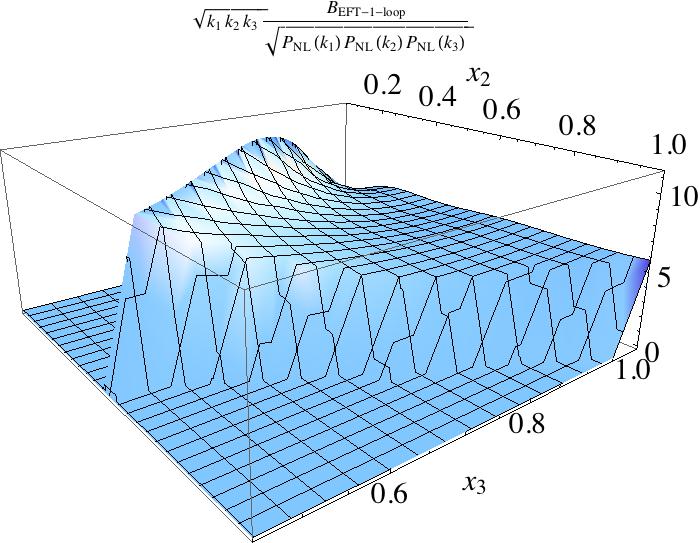}
\includegraphics[width=0.41\textwidth]{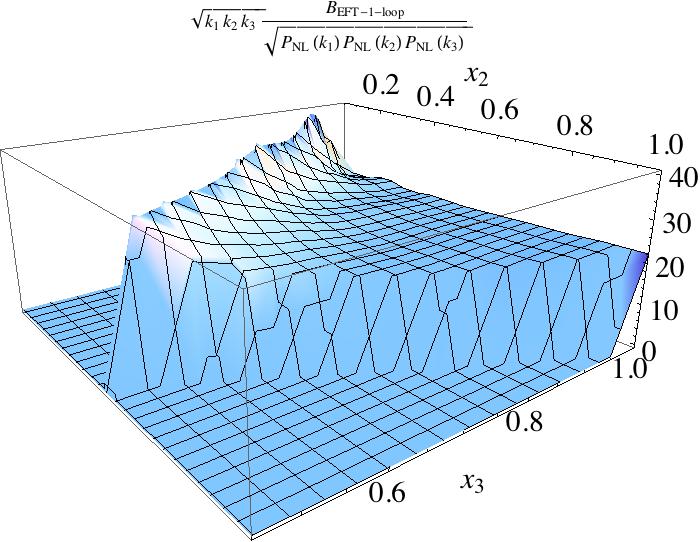}
\caption{ \label{fig:shape-plots-final} \footnotesize\it  The shape of the dark matter bispectrum given by the EFTofLSS, with $k_1=0.1\hinvMpc$ on the left, and $k_1=0.3\hinvMpc$ on the right.}
\end{center}
\end{figure}

In Fig.~\ref{fig:pval-kmax}, we display the $p$-values corresponding the one-loop EFT prediction,
\beq
B_\text{EFT-1-loop} = B_\text{tree}+B_\text{1-loop}+B_{c_{\rm s}}\ ,
\eeq
along with the one-loop SPT and tree-level predictions, as a function of $k_{\rm comp}$ in the top panel and $k_{\rm comp}^3$ in the bottom panel. In the middle panel, we also show the reduced $\chi^2$ for these predictions. Once $\co$ is fixed by fitting to the matter power spectrum, the one-loop EFT calculation has {\em no} free parameters and extends the range of the prediction compared to one-loop SPT by $\sim$65\%, which corresponds to a factor of $\sim$4.5 more modes whose behavior is reliably captured by the theory~\footnote{Notice that the error bars on the triangles are larger at low $k$'s that at high $k$'s, which means that the improvement of SPT with respect to linear theory might be a somewhat overestimated}. The bottom panel of Fig.~\ref{fig:pval-kmax} gives a pictorial representation of this improvement. We summarize the shape of the EFT bispectrum for dark matter, that is valid up to $k\simeq 0.25\hinvMpc$, in Fig.~\ref{fig:shape-plots-final}.

\begin{figure}[t]
\begin{center}
\includegraphics[width=0.9\textwidth]{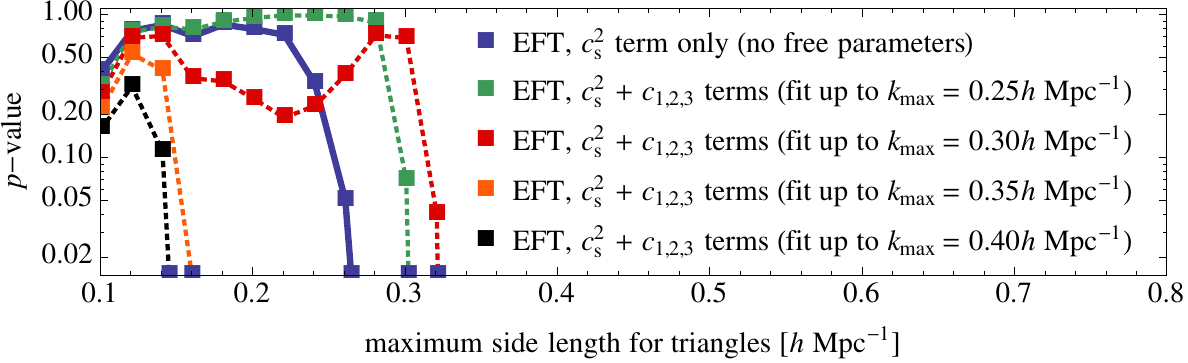}\\[0.3cm]
\includegraphics[width=0.9\textwidth]{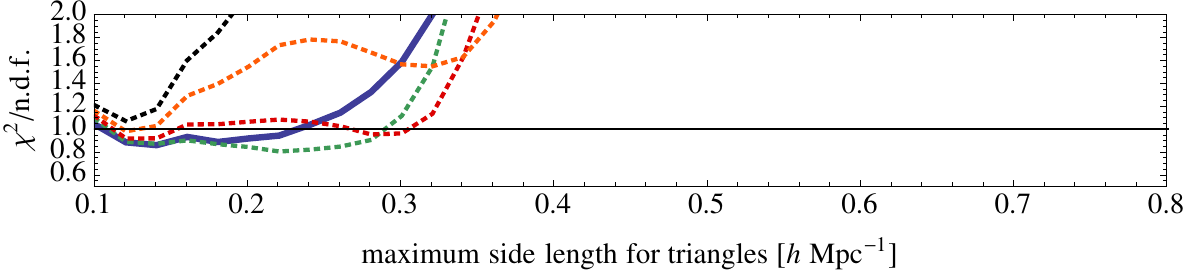}
\caption{ \label{fig:pval-kmax-c123}  \footnotesize\it  
As Fig.~\ref{fig:pval-kmax}, but showing curves corresponding to the one-loop EFT prediction with the 3 additional counterterms from Sec.~\ref{sec:counterterms} included, with their coefficients fit to nonlinear bispectrum data up to the maximum $k$ indicated in the legend. We find that these 3 terms can only slightly improve the reach of the prediction, indicating that a higher-order contribution (e.g.\ $B_\text{2-loop}$) is important there and therefore that these 3 terms can safely be neglected.}
\end{center}
\end{figure}

As a check that we are not neglecting any important terms, we also repeat this analysis for the sum of $B_\text{EFT-1-loop}$ plus the 3 tree-level counterterms we have mentioned earlier:
\beq
B_\text{EFT-1-loop+3} = B_\text{EFT-1-loop}+B_{c_1}+B_{c_2}+B_{c_3}\ .
\eeq
We simultaneously fit for the values of $c_1$, $c_2$, and $c_3$ (while keeping $\co$ fixed), using all triangles with maximum side length less than some upper limit $k_{\rm max}$. The results for various values of $k_{\rm max}$ are shown in Fig.~\ref{fig:pval-kmax-c123}. We find that there is no issue with regard to ``over-fitting": even with the freedom of 3 arbitrary parameters, the fit cannot reach beyond $k_{\rm comp}\sim 0.32 \hinvMpc$ before becoming much worse at low~$k$. Therefore, in this region there must be a higher-order correction, namely $B_\text{2-loop}$ and associated counterterms, that contributes significantly to the nonlinear bispectrum.

Furthermore, by examining Fig.~\ref{fig:pval-kmax-c123}, we find that the 3 extra counterterms are not able to improve the reach of the prediction into the UV more than about $\Delta k\sim 0.07 \hinvMpc$  as compared to the $B_\text{EFT-1-loop}$ curve.  The ratio of $B_\text{2-loop}$ and $B_\text{1-loop}$ is expected to scale roughly like $k/\knl$ in this range, implying that if $B_\text{2-loop}$ is significant at $k\sim 0.32 \hinvMpc$, it will also be significant at $k\sim 0.25\hinvMpc$, where $B_\text{EFT-1-loop}$ fails. Therefore, there is no regime in which the 3 extra counterterms are more important than $B_\text{2-loop}$, and so, as anticipated by our earlier estimates, it is self-consistent to neglect these 3 terms (we must still include~$B_{c_{\rm s}}$, though, because it is enhanced relative to the other counterterms by a factor of~$2\pi$)~\footnote{%
This interpretation is further confirmed by the fact that the best fit values for the $c_{1,2,3}$ parameters giving rise to the red dotted line in Fig.~\ref{fig:pval-kmax-c123} (which corresponds to a fit using triangles with $k_{\rm max}\leq 0.3\invMpc$) are respectively equal to $c_1=1.26\,{\rm Mpc}^2/h^2,\;c_2=-3.68\,{\rm Mpc}^2/h^2,\;c_3=9.11\,{\rm Mpc}^2/h^2$. An equally good fit is obtained just using the $B_{c_2}$ and $B_{c_3}$ counterterms, whose have best fit values are $c_2= 0.47\,{\rm Mpc}^2/h^2,\;c_3=5.48\,{\rm Mpc}^2/h^2$. Some of these numbers are quite larger than naively expected, even after considering the small normalization of the shapes of $B_{c_2}$ and $B_{c_3}$. This further suggests that these counterterms are trying to compensate for the lack  in the theoretical prediction of the  two-loop diagram, which is important in that $k$-range.  Of course it would be nice to explicitly and more definitively verify this argument with an higher order calculation, something that we leave to future work.
}. 

\begin{figure}[t]
\begin{center}
\includegraphics[width=0.95\textwidth]{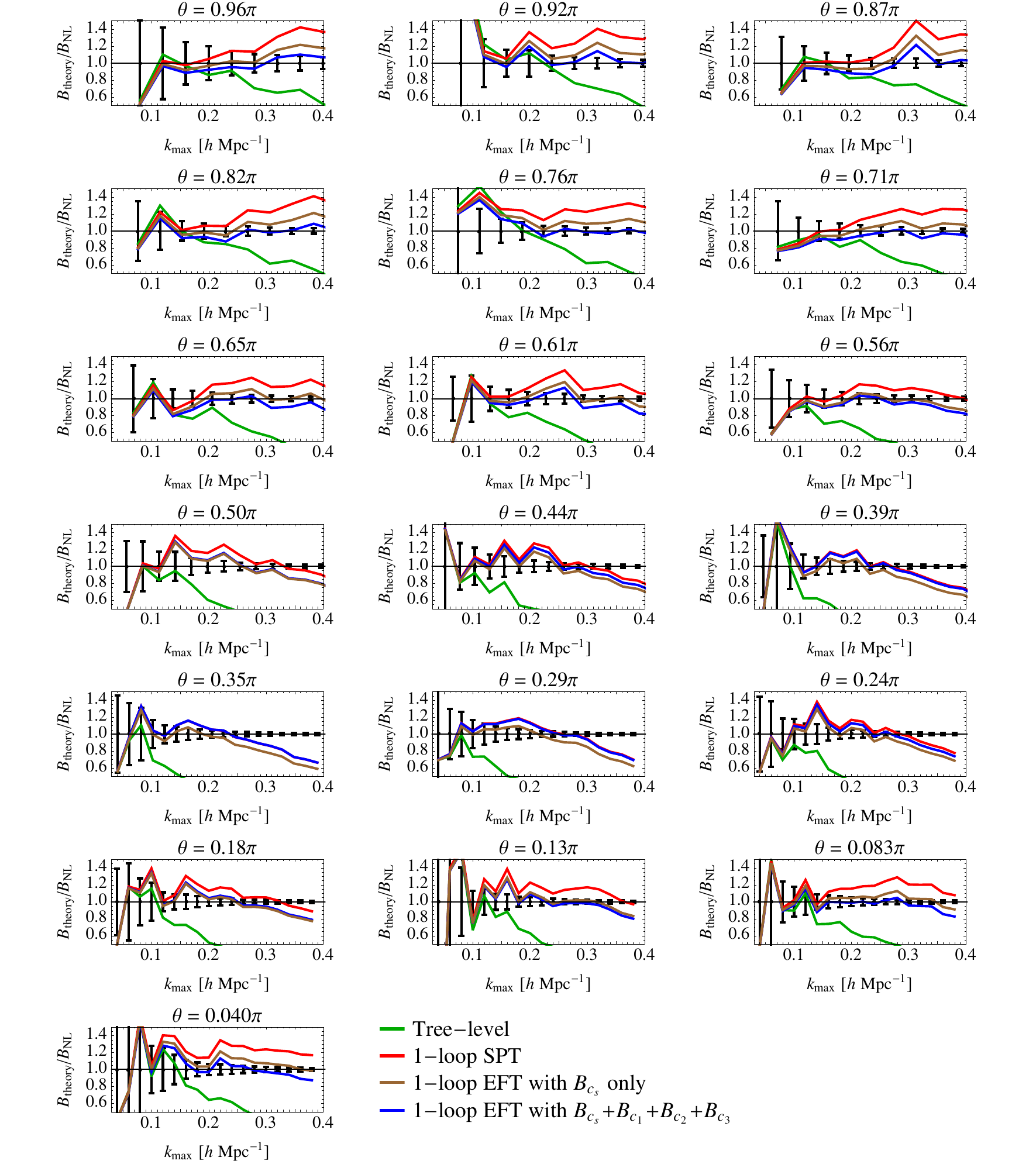}
\caption{ \label{fig:fits-isos} \footnotesize\it   Comparison of various theory curves, normalized to nonlinear bispectrum data for isosceles triangles ($k_2=k_1$). The blue curve is fit using all triangles with maximum side length less than $0.3\hinvMpc$. The EFT prediction fails at lower $k$ on equilateral configurations than on flat or squeezed triangles, but on average, agreement with the errorbars is obtained for $k\lesssim 0.25\hinvMpc$, even when no free parameters are fit to the bispectrum data.}
\end{center}
\end{figure}

\begin{figure}[t]
\begin{center}
\includegraphics[width=0.95\textwidth]{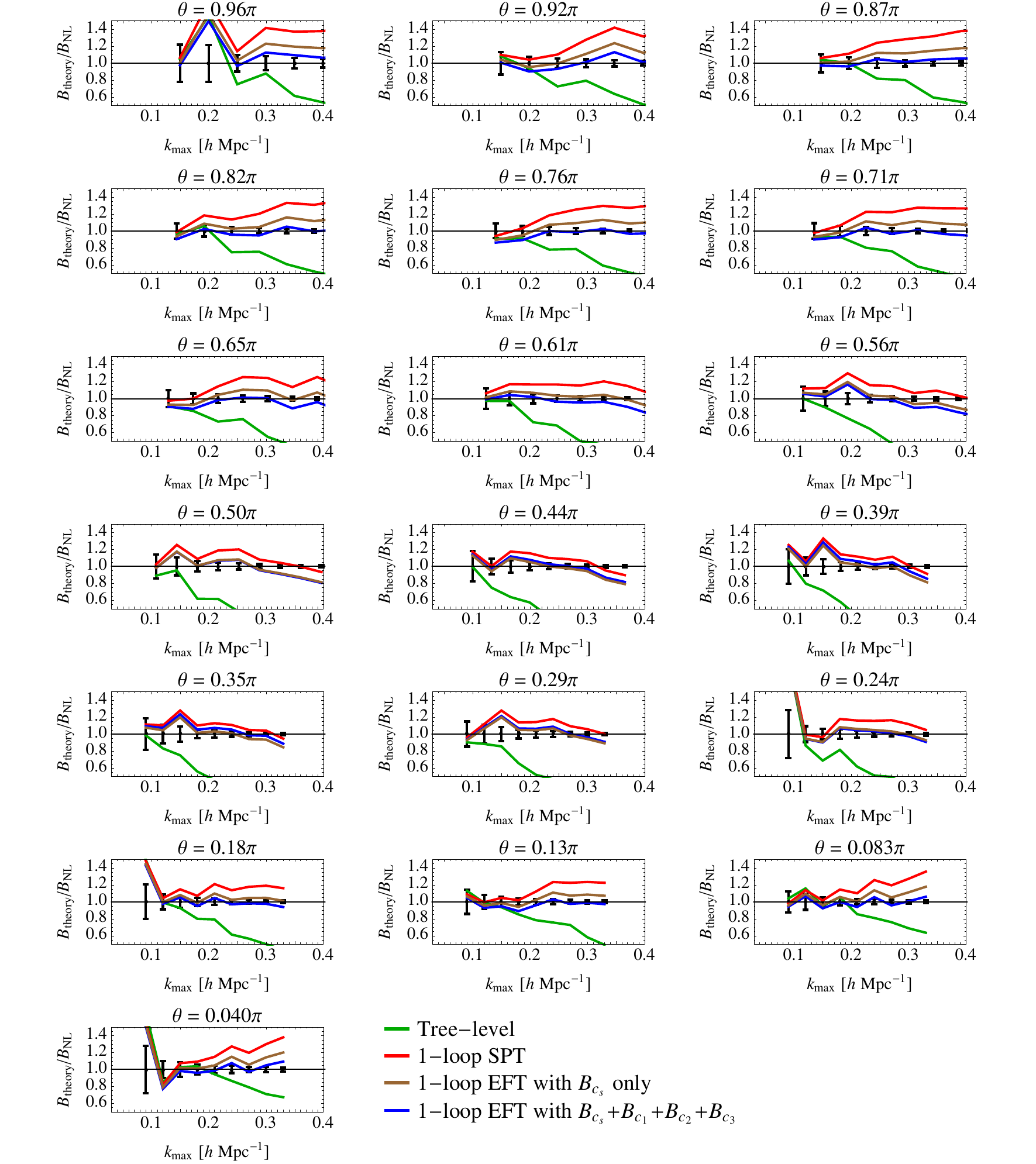}
\caption{ \label{fig:fits-3o2} \footnotesize\it   As Fig.~\ref{fig:fits-isos}, but for triangles with $k_2=1.5k_1$.}
\end{center}
\end{figure}

\begin{figure}[t]
\begin{center}
\includegraphics[width=0.95\textwidth]{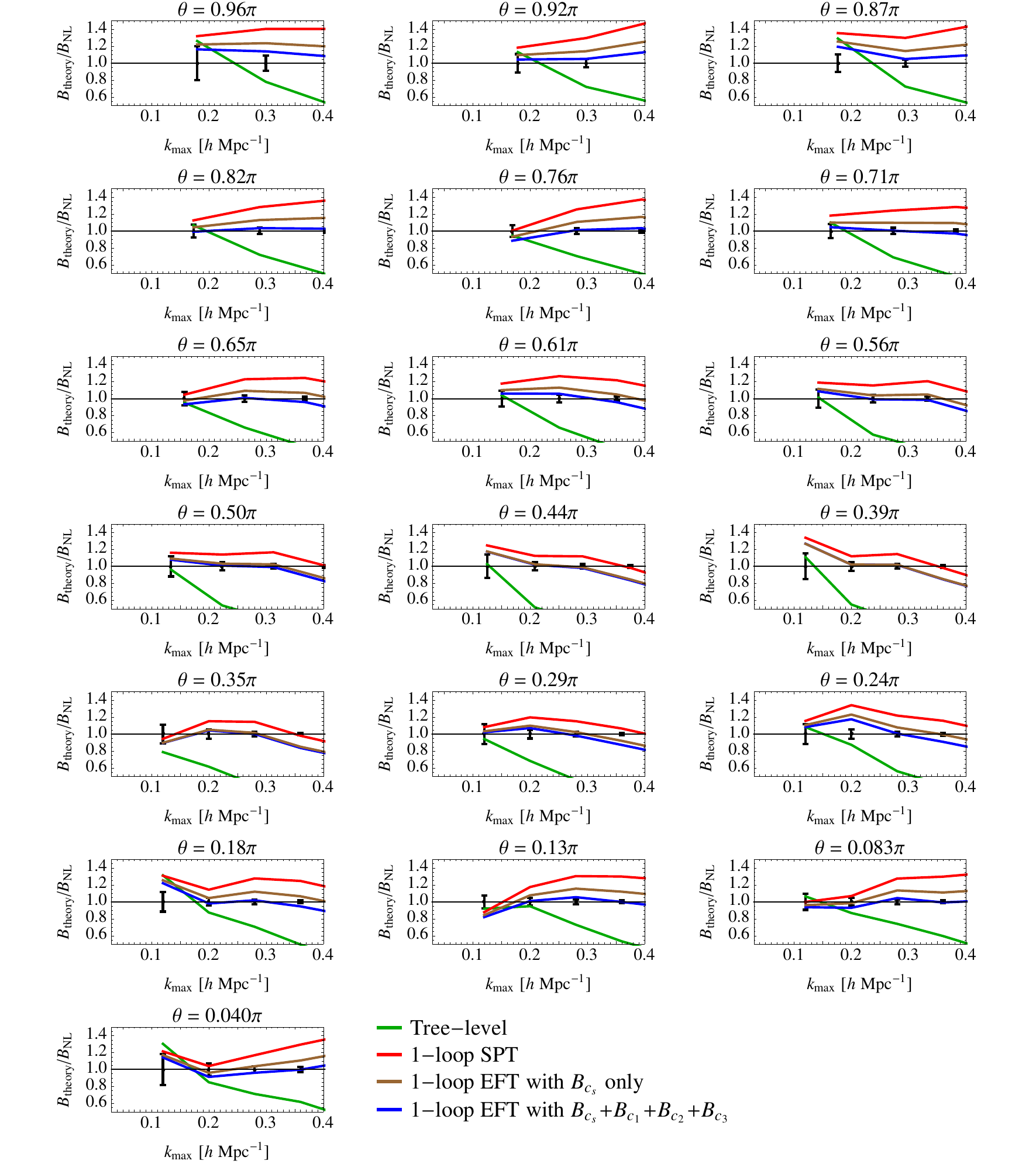}
\caption{ \label{fig:fits-2} \footnotesize\it  As Fig.~\ref{fig:fits-isos}, but for triangles with $k_2=2k_1$.}
\end{center}
\end{figure}

More in detail, Figures~\ref{fig:fits-isos},~\ref{fig:fits-3o2}, and~\ref{fig:fits-2} display comparisons of various theory curves with the nonlinear data. There is some configuration-dependence in the maximum wavenumber up to which the EFT prediction can reliably reach, but the average limit of validity is around $k\sim0.25\hinvMpc$, similar to the reach of the one-loop prediction for the matter power spectrum with similar error bars. In addition, for equilateral triangles  $B_\text{EFT-1-loop}$ deviates from the data by around 20\% at $k\sim 0.30\hinvMpc$, as predicted (within $\mathcal{O}(1)$) by our prior estimate for $B_{332}$.

To compute the EFT predictions, we have used CAMB~\cite{Lewis:1999bs} to generate a linear power spectrum, and a version of the Copter code~\cite{Carlson:2009it} modified to utilize the IR-safe integrands of~\cite{Carrasco:2013sva} and Monte Carlo integration routines from the CUBA library~\cite{CUBA}, to compute $P_\text{1-loop}$ and $B_\text{1-loop}$.

\section{Conclusions}

In this paper we have studied the prediction for the equal-time bispectrum of dark matter at redshift $z=0$ at one loop in the EFTofLSS. Since the equal time matter bispectrum is IR-safe, we have contented ourselves with not performing the IR-resummation of~\cite{Senatore:2014via}, which leaves us with small residual $\sim2\%$ oscillations that we will address in future work. Our main interest here was to study the UV reach of the prediction of the EFTofLSS. At one loop, the power spectrum at $z=0$ agrees with numerical simulations at about percent level up to $k\simeq 0.3\hinvMpc$~\cite{Carrasco:2012cv} at one loop\footnote{%
However, the reach of the one-loop prediction must always be specified along with the precision that is requested of the match between the prediction and nonlinear data.
} and up to $k\simeq 0.6\hinvMpc$ at two loops~\cite{Carrasco:2013mua}, after the IR-resummation has been performed~\cite{Senatore:2014via}. This prediction requires us to fix a parameter, the so called speed of sound $c_s^2$, to account for the mistakes from the short distance physics that are accidentally included when doing perturbative calculations. When passing to the bispectrum, we have found that there is no need to include any additional parameter when performing the calculation at one loop. This is so because, at the relevant wavenumbers, the additional counterterms that could be inserted in the bispectrum calculation will contribute in a way that is not safely larger than the two-loop contribution that we do not compute. The only counterterm that we are justified to insert is the $c_s$ term that had already appeared in the power spectrum calculation and had been measured from that observable. For this reason, without any additional parameter to fit, we find that the prediction of the EFTofLSS for the equal-time bispectrum of dark matter at $z=0$ agrees very well with $N$-body simulations up to $k\simeq 0.25\hinvMpc$, given the accuracy of the measurements, which is of order a few percent at the highest $k$ of interest~\footnote{%
The error is dominated by the difference of the power spectrum used in the EFT calculation, where we used the realizations-averaged power spectrum, and the actual $\delta$'s simulated by the numerical code.
}.  This is a factor of 65\% better than SPT~\footnote{%
The improvement at one loop with respect to SPT is less strong than in the power spectrum. This might be due to the fact that the relative weight of the counterterms with respect to the loops is diminished as we increase the number of external legs, or to the fact that the errors in the bispectrum measurement are larger at low wavenumbers where SPT fails.
}. While the fit is very good on average up to  $k\simeq 0.25\hinvMpc$, the fit performs somewhat worse on equilateral configurations, in agreement with expectations that for a given maximum $k$, equilateral triangles are more nonlinear than for the other shapes. We should also point out that the simulation we use has a $\sigma_8$ value which is about 10\% higher than the current preferred value for our cosmology, so it is expectable that the comparison with observations or simulations with lower $\sigma_8$ or at higher redshifts will show an even better agreement. Still, given the large number of triangular configurations, the average reach of validity of the fit is $k\simeq 0.25\hinvMpc$.

This is a very promising result. The scale up to which we are able to match the data is comparable to the one up to which the EFTofLSS matches numerical data in the power spectrum at one loop. This is an important confirmation of the paradigm of the EFTofLSS, because it is expected that all quantities, when evaluated at the same order in perturbation theory, should match the numerical data to approximately the same wavenumber.

Additionally, it is quite remarkable that we can achieve our result without the addition of a new parameter. The value of $c_s^2$ is fitted to the power spectrum, and the prediction for the bispectrum follows without any freedom. In fact, we can even ask if the fit can be improved by adding additional quadratic counterterms to the calculation. This is a useful check because there is some uncertainty on the actual size of the various terms, so that it is possible that the additional quadratic counterterms might potentially contribute in a larger way than the two-loop contribution. We instead find that after adding the additional counterterms, the $k$-reach of the fit is  improved only slightly, which shows that the two-loop contribution is relevant roughly at the scales where our one-loop calculation fails. Somewhat contrary to what happens for the power spectrum, it is very straightforward to identify the $k$-value when the prediction of the EFT stops matching the data. Finally, we notice that the same $c_s^2$ parameter determined from the matter power spectrum can predict the momentum power spectrum so that it agrees with the numerical data up to $k\simeq 0.3\invMpc$. In other words, the two-loop matter power spectrum, the one-loop momentum power spectrum, and now the one-loop bispectrum are all predicted with only one parameter.

There are a number of ways that we envision to proceed. First, it would be interesting to have more accurate $N$-body simulations, to tighten even more the error bars and understand better the performance of the prediction. Alternatively, and potentially  even better, we could perform the EFT calculations directly for the actual realization being simulated by the codes, so that cosmic variance would be made negligible. Second, as it was done for the power spectrum, we would like to extend the calculation to two loops and include IR-resummation. These results obtained in the context of the EFTofLSS keep suggesting that there is a much larger number of cosmological modes that are amenable to analytic techniques. We believe this program is very important to assess what we will learn on primordial cosmology in the next decade, and therefore we  are eager to  push it forward.

\section*{Acknowledgments}
Raul E. Angulo and Marcel Schmittfull provided the measurement from the numerical simulations, while the other authors performed the analytic calculations in the EFT and the fits to the data.

We thank Simon White for a brief but intense conversation that helped motivating this project. We thank Daniel Green for initial collaboration. We thank Oliver Hahn, Eiichiro  Komatsu,  Uro\v{s} Seljak, Rashid Sunyaev and Matias Zaldarriaga for interesting conversations.   S.F. is partially supported by the Natural Sciences and Engineering Research Council of Canada.  L.S. is supported by DOE Early Career Award DE-FG02-12ER41854 and by NSF grant PHY-1068380.

\appendix
\section*{Appendix}
\section{Solving the Equations of Motion}
\label{sec:eftreview}

In this appendix, we review the method to perturbatively solve the equations of motion for $\delta$ and $v^i$ in the EFTofLSS, using the approximation that the time-dependence of each solution is given by an appropriate power of the linear growth factor $D_1(a)$ (normalized to unity at $a=1$). If we wish to only make predictions for $z=0$, it is acceptable to approximate the time-dependence in this way, since the error incurred by this approximation has been shown numerically to be very small~\cite{Carrasco:2012cv}.\footnote{%
A more thorough investigation of this point, as well as a formalism that allows computations to be carried out using the full time-dependence, can be found in~\cite{exacttimedep}.
} In addition, since we only work up to one-loop order, any errors made by this approximation can be partially absorbed into the free parameters we fit to data.

\subsection{Loop corrections and linear counterterms}
\label{sec:eftreview-linear}

We first review the solutions arising from the standard nonlinear terms ($\sim\delta\theta$ and $\sim\theta\theta$, where $\theta \equiv \d_i v^i$) appearing in the equations of motion, as well as from the single linear counterterm ($\sim\d^2\phi$) that has previously been considered in studies of the EFTofLSS, as it is the relevant counterterm for the matter power spectrum at one and two loops~\cite{Carrasco:2012cv,Carrasco:2013mua}. We refer the reader to~\cite{Carrasco:2013mua,Carrasco:2012cv} for a more extensive discussion of the motivations and assumptions underlying the formalism we present below.

The starting point is the equations of motion from~\cite{Carrasco:2013mua}, including only the term from the stress tensor that is linear in $\d^2\phi$, integrated against an unknown kernel that parametrizes the non-locality in time of the coupling between long and short modes of the fields:
\bea
\label{eq:master2}
\nonumber
&&a\H \delta'+\theta= - \! \int_{\vq} \alpha(\vkp,\vk-\vkp)\delta(\vk-\vkp)\theta(\vkp)\ , \\
\nonumber
&&  a\H \theta'+\H \theta+\frac{3}{2} \H_0^2 \, \Omm  \frac{a_0^3}{a} \delta
= - \! \int_{\vq} \beta(\vkp,\vkkp)\theta(\vk-\vkp)\theta(\vkp) \\
&&\qquad\qquad\qquad\qquad\qquad\qquad\quad
+ \, \epsilon k^2 \int \frac{da'}{a'\H(a')}   \kappa_1(a,a')\; [ \d^2\phi(\tau',\vx_{\rm fl})]_{\vk}\ ,  
\eea
where $\epsilon$ is a parameter inserted to organize the powers of $\kappa_1(a,a')$ appearing in the solutions we will derive, and 
where
\be
\alpha(\vk,\vkp)=\frac{\left(\vk+\vkp\right)\cdot\vk}{k^2}\ ,\qquad\beta(\vk,\vkp)=\frac{\left(\vk+\vkp\right)^2 \vk\cdot\vkp}{2 q^2 k^2}\ .
\ee
We use the labelled brackets $ [ f(\vx_{\rm fl}(\vx))]_{\vk}$ to mean that we take the Fourier transform of the given function $f$ and evaluate it at the momentum~$\vk$: $[f(\vx_{\rm fl}(\vx))]_{\vk}\equiv\int d^3x\; e^{- i\vx\cdot\vk} f(\vx_{\rm fl}(\vx))$. We use $\vx_{\rm fl} \equiv \vx_{\rm fl}[\tau,\tau'] = \vx - \int_{\tau'}^\tau d\tau'' \vv(\tau'',\vx_{\rm fl}[\tau,\tau''])$ as the argument of $\phi$ to ensure that the equations of motion are diffeomorphism-invariant, and also because the behavior of a mode along its entire past trajectory could in principle have an influence on other modes. However, since we will only consider up to one-loop counterterms in this work, and the difference between $\vx_{\rm fl}$ and $\vx$ only becomes apparent in two- or higher-loop calculations, we can take $\vx_{\rm fl}\approx\vx$. We can also use Poisson's equation to rewrite $\d^2\phi$ in terms of $\delta$, transforming the second equation into
\bea
\nn
&& a\H \theta'+\H \theta+\frac{3}{2} \H_0^2 \, \Omm  \frac{a_0^3}{a} \delta
= - \! \int_{\vq} \beta(\vkp,\vkkp)\theta(\vk-\vkp)\theta(\vkp) \\
&&\qquad\qquad\qquad\qquad\qquad\qquad\quad
+ \, \epsilon  k^2 \int \frac{da'}{a'\H(a')} K(a,a') \delta(a',\vk) \ ,
\eea
where
\beq
K(a,a') \equiv \frac{3}{2} H_0^2 \, \Omm \frac{a_0^3}{a'}\, \kappa_1(a,a') \ .
\eeq

These equations can be solved using the following ansatz:
\bea
\label{eq:deltaexp}
\delta(a,\vk) &=& \sum_{n=1}^\infty \, [D_1(a)]^n \delta^{(n)}(\vk)
+ \epsilon \sum_{n=1}^\infty \, [D_1(a)]^{n+\zeta} \tilde{\delta}^{(n)}(\vk)\ , \\
\label{eq:thetaexp}
\theta(a,\vk) &=& -\H(a) f \sum_{n=1}^\infty \, [D_1(a)]^n \theta^{(n)}(\vk)
-\epsilon \H(a) f \sum_{n=1}^\infty \, [D_1(a)]^{n+\zeta} \tilde{\theta}^{(n)}(\vk)\ ,
\eea
under the assumptions that $\delta^{(n)}(a',\vk) = [D_1(a')/D_1(a)]^n \, \delta^{(n)}(a,\vk)$ and $\Omm(a)\approx f^2$, where $f \equiv \partial \log D_1/\partial \log a$. (Recall that these assumptions allow us to solve for the momentum-dependent part of the solution at each order using recurrence relations that do not involve the time variable, as explained in~\cite{Carrasco:2013mua}.) The $\vk$-dependent parts of the solutions are written in terms of kernels $F_n$ and $G_n$ in the following manner:
\bea
\label{eq:deltakansatz}
\delta^{(n)}(\vk) &=& \int_{\vq_1} \cdots \int_{\vq_n}
(2\pi)^3 \delta_{\rm D}(\vk-\vq_{1\cdots n})
F_n(\vq_1,\dots,\vq_n)
\delta^{(1)}(\vq_1)\cdots\delta^{(1)}(\vq_n)\ , \\
\theta^{(n)}(\vk) &=&  \int_{\vq_1} \cdots \int_{\vq_n}
(2\pi)^3 \delta_{\rm D}(\vk-\vq_{1\cdots n})
G_n(\vq_1,\dots,\vq_n)
\delta^{(1)}(\vq_1)\cdots\delta^{(1)}(\vq_n)\ ,
\label{eq:thetakansatz}
\eea
with $\tilde{\delta}^{(n)}(\vk)$ and $\tilde{\theta}^{(n)}(\vk)$ written analogously in terms of $\tilde{F}_n$ and $\tilde{G}_n$. Once $F_n$, $G_n$, $\tilde{F}_n$, and $\tilde{G}_n$ are specified, then any correlation function of $\delta$ or $\theta$ can be calculated up to a specified order. The resulting expression will be in terms of the linear matter power spectrum, defined by
\beq
\left\la \delta^{(1)}(\vk) \delta^{(1)}(\vk') \right\ra = (2\pi)^3 \delta_{\rm D}(\vk+\vk') P_{11}(k)\ ,
\eeq
which is calculated numerically by a Boltzmann code such as CAMB~\cite{Lewis:1999bs} and used as an input to the EFTofLSS computation.

Plugging Eqs.~(\ref{eq:deltaexp}) to~(\ref{eq:thetakansatz}) into the equations of motion and collecting terms of order $\epsilon^0$ yields the following recurrence relations for $F_n$ and $G_n$, familiar from SPT (e.g.~\cite{Bernardeau:2001qr}):
\bea
\nn
F_n(\vq_1,\dots,\vq_n) &=& \sum_{m=1}^{n-1} \frac{G_m(\vq_1,\dots,\vq_m)}{(2n+3)(n-1)}
\left[ (2n+1) \frac{\vk\cdot\vk_1}{k_1^2} F_{n-m}(\vq_{m+1},\dots,\vq_n) \right. \\
\nn
&&\qquad\qquad\qquad\qquad\qquad \left. +  \frac{k^2 (\vk_1\cdot\vk_2)}{k_1^2 k_2^2}
G_{n-m}(\vq_{m+1},\dots,\vq_n) \right], \\
\nn
G_n(\vq_1,\dots,\vq_n) &=& \sum_{m=1}^{n-1} \frac{G_m(\vq_1,\dots,\vq_m)}{(2n+3)(n-1)}
\left[ 3 \frac{\vk\cdot\vk_1}{k_1^2} F_{n-m}(\vq_{m+1},\dots,\vq_n) \right. \\
&&\qquad\qquad\qquad\qquad\qquad \left. + n \frac{k^2 (\vk_1\cdot\vk_2)}{k_1^2 k_2^2}
G_{n-m}(\vq_{m+1},\dots,\vq_n) \right]\ .
\label{eq:sptrecurrence}
\eea
Meanwhile, the terms of order $\epsilon^1$ (corresponding to solutions involving a single power of $K(a,a')$) give recurrence relations for $\tilde{F}_n$ and $\tilde{G}_n$, under the assumption that
\beq
\label{eq:cnassumption}
c_n(a)  = \bar c_n ( \xi D_1(a)^\zeta \H^2 f^2) \ ,
\eeq
where
\beq
c_n(a) \equiv \int \frac{da'}{a'\H(a')} K(a,a') \frac{D_1(a')^n}{D_1(a)^n}
\eeq
and $\xi$ and $\zeta$ are constants. In~\cite{Carrasco:2013mua}, the values $\zeta=2$ and $\xi=9$ were used; this choice for $\zeta$ arises from the assumption that the coefficient of the $k^2 P_{11}$ counterterm in the matter power spectrum has the same time-dependence as $P_\text{1-loop}$, while the choice for $\xi$ cancels factors of $1/9$ that would otherwise have appeared in  the $\tilde{F}_n$ kernels. (Note that changing $\xi$ simply redefines $\bar{c}_n$, so we are free to set $\xi$ to a convenient value.) In Eq.~(\ref{eq:cnassumption}), leaving $\zeta$ arbitrary allows for a more general time-dependence of the $k^2 P_{11}$ counterterm in the power spectrum, as discussed further in~\cite{Foreman:2015uva}.\footnote{%
In a cosmological model in which loop corrections diverge (for example, a scaling universe with sufficiently high tilt), recall that the $k^2 P_{11}$ counterterm in the power spectrum will have two parts: one that depends on the method used to regulate the loop integrals, whose role is to cancel the regulator-dependence (e.g.~$\Lambda$-dependence in the case of a hard momentum cutoff $\Lambda$) of the result, and another whose role is to capture the finite effects of the UV modes that have been ``integrated out" of the theory. In the formalism presented here, this would be incorporated by splitting $K(a,a')$ into two separate functions $K_\Lambda(a,a')$ and $K_\text{finite}(a,a')$, each associated with a different value of $\zeta$ and different perturbative solutions. For the realistic cosmology we consider in this paper, however, the loop corrections converge on their own and do not require a regulator, and so we only need to consider $K(a,a')$ and $\zeta$ corresponding to the finite counterterm.
} The resulting recurrence relations are the following:
\begin{align}
\nn
\tilde{F}_n(\vq_1,\dots,\vq_n) &= \frac{1}{(n-1+\zeta)(n+\frac{3}{2}+\zeta)} 
	\times \left\{ -\xi \bar{c_n} \, k^2 F_n(\vq_1,\dots,\vq_n)  \right. \\
&\quad + \left. \sum_{m=1}^{n-1} \lb (n+\tfrac{1}{2}+\zeta) \alpha(\vk_1,\vk_2)
\mathcal{A}_m(\vq_1,\dots,\vq_n) 
	+ \beta(\vk_1,\vk_2) \mathcal{B}_m(\vq_1,\dots,\vq_n)\rb  \right\}\ , \\
\nn
\tilde{G}_n(\vq_1,\dots,\vq_n) &= \frac{1}{(n-1+\zeta)(n+\frac{3}{2}+\zeta)} 
	\times \left\{ -\xi (n+\zeta) \bar{c_n} \, k^2 F_n(\vq_1,\dots,\vq_n)  \right.  \\
&\quad + \left. \sum_{m=1}^{n-1} \lb \tfrac{3}{2} \alpha(\vk_1,\vk_2)
\mathcal{A}_m(\vq_1,\dots,\vq_n) 
	+ (n+\zeta) \beta(\vk_1,\vk_2) \mathcal{B}_m(\vq_1,\dots,\vq_n) \rb \right\}\ ,
\end{align}
where
\bea
\nn
&&\mathcal{A}_m(\vq_1,\cdots\,\vq_n) \equiv \\
&&\qquad \tilde{G}_m(\vq_1,\dots,\vq_m)  F_{n-m}(\vq_{m+1},\dots,\vq_n)
+ G_m(\vq_1,\dots,\vq_m)  \tilde{F}_{n-m}(\vq_{m+1},\dots,\vq_n) \ , \\
\nn
&& \mathcal{B}_m(\vq_1,\cdots\,\vq_n) \equiv \\
&&\qquad \tilde{G}_m(\vq_1,\dots,\vq_m)  G_{n-m}(\vq_{m+1},\dots,\vq_n)
+ G_m(\vq_1,\dots,\vq_m)  \tilde{G}_{n-m}(\vq_{m+1},\dots,\vq_n) \ .
\eea

The value of $\xi$ can be fixed by examining the lowest-order counterterm appearing in the matter power spectrum,
\beq
\left\la \tilde{\delta}^{(1)}(\vk) \delta^{(1)}(\vk') \right\ra 
	= (2\pi)^3 \delta_{\rm D}(\vk+\vk') P_\text{tree}^{(c_{\rm s})}(k)\ ,
\eeq
which involves $\tilde{F}_1(\vk)$:
\beq
\tilde{F}_1(\vk) = -\frac{\xi}{\zeta(\zeta+\frac{5}{2})} \bar{c}_1 k^2\ .
\eeq
To simplify this expression, we set $\xi=\zeta(\zeta+5/2)$, implying that
\beq
P_\text{tree}^{(c_{\rm s})}(k,a) = -2 [D_1(a)]^{2+\zeta} \bar{c}_1 k^2 P_{11}(k)\ .
\eeq
This explicitly shows the relationship between $\zeta$ and the time-dependence of the tree-level counterterm in the power spectrum.

Now, to fix the value of $\zeta$, one can utilize the fact that over the range where we expect the one-loop EFT prediction for the power spectrum to be valid, $0.10\hinvMpc \lesssim k \lesssim 0.30\hinvMpc$, the linear power spectrum resembles that of a scaling universe with tilt $n\approx-1.7$. In such a universe, the equations of motion obey a scaling symmetry that uniquely determines the value of $\zeta$ to be $4/(3+n)\approx3.1$, and this is also approximately the value that results from a fit of the one-loop EFT prediction to nonlinear power spectra at redshifts $z\lesssim 1$~\cite{Foreman:2015uva}. Therefore, in this paper we use $\zeta=3.1$ in our numerical calculations. Also, dimensional analysis and the presence of a trivial angular integral in $P_\text{1-loop}$ determine the following relationship between $\bar{c}_1$ and the parameter we ultimately fit for, $\co$:
\beq\label{eq:cbartocs}
\bar{c}_1 = \frac{(2\pi)\co}{\knl^2}\ .
\eeq

As mentioned in Sec.~\ref{sec:counterterms}, the $\tilde{\delta}$ solutions contribute to two counterterms (displayed in Eq.~(\ref{eq:bcsdef}) for $z=0$) to the bispectrum, arising from the correlators of the form $\langle \tilde{\delta}^{(2)}(\vk_1) \delta^{(1)}(\vk_2) \delta^{(1)}(\vk_3)\rangle$ and $\langle \tilde{\delta}^{(1)}(\vk_1) \delta^{(2)}(\vk_2) \delta^{(1)}(\vk_3)\rangle$ (plus permutations of external momenta). The second one involves $\tilde{F}_1$ and $F_2^{\rm (s)}$, while the first one involves $\tilde{F}_2^{\rm (s)}$, which takes the form
\begin{align}
\nn
&-(1+\zeta)(7+2\zeta) \tilde{F}_2^{\rm (s)}(\vk_1,\vk_2) =
	\bar{c}_1 \left\{ \lp 5 + \frac{113\zeta}{14} + \frac{17\zeta^2}{7} \rp (k_1^2+k_2^2)
	+ \lp 7 + \frac{148\zeta}{7} + \frac{48\zeta^2}{7} \rp \vk_1\cdot\vk_2 \right. \\
\nn
&\qquad\qquad + \lp 2+\frac{59\zeta}{7}+\frac{18\zeta^2}{7} \rp 
	\lp \frac{1}{k_1^2}+\frac{1}{k_2^2} \rp  (\vk_1\cdot\vk_2)^2 
+ \lp \frac{7}{2}+\frac{9\zeta}{2}+\zeta^2 \rp \lp \frac{k_1^2}{k_2^2} + \frac{k_2^2}{k_1^2} \rp
	\vk_1\cdot\vk_2 \\
&\qquad\qquad \left. + \lp \frac{20\zeta}{7} + \frac{8\zeta^2}{7} \rp 
	\frac{(\vk_1\cdot\vk_2)^3}{k_1^2 k_2^2} \right\}
\label{eq:f2stilde}
\end{align}
in the local-in-time limit of $K(a,a')$ (for which $\bar{c}_n=\bar{c}_1$ for all $n$). In the nonlocal case, extra factors of $\bar{c}_2$ will appear in $\tilde{F}_2^{\rm (s)}(\vk_1,\vk_2)$, but unless the nonlocality is very severe, $\bar{c}_2$ will be very close to $\bar{c}_1$ (see~\cite{Carrasco:2013mua} for a more precise statement, involving parameterizing the nonlocality by the power of the growth factor $D_1(a')$ appearing in $K(a,a')$).

\subsection{Quadratic counterterms}

As mentioned in Sec.~\ref{sec:counterterms}, the linear term in $\gammai{}^i$ does not exhaust the possible tree-level counterterms that can enter into the bispectrum calculation---there are also three independent quadratic terms consistent with symmetries that can be written down, along with one linear combination of fields that only contributes at quadratic and higher order:
\begin{align}
\d_i \gammai{}^i &\supset
 \d^2 \delta^2 + 
	\d^2 \lb \frac{ \d^j \d^k}{\d^2} \delta \cdot  \frac{\d_j \d_k}{\d^2}\delta \rb +
	\d_i \lb \frac{\d^i\d^j }{\d^2} \delta \cdot \d_j \delta \rb +
	\d^2 \lb \frac{\theta}{-\H(a) f} - \delta \rb\ .
\label{eq:dstress4cts}
\end{align}
Taking the Fourier transform of each term and inserting into the Euler equation in the same way as for the linear term, we obtain the following:
\begin{align} \nn
&a\H \theta'+\H \theta+\frac{3}{2} \H_0^2 \, \Omm  \frac{a_0^3}{a} \delta \\ \nn
&\quad = - \! \int_{\vq} \beta(\vkp,\vkkp)\theta(\vk-\vkp)\theta(\vkp)
	+ k^2  \int \frac{da'}{a'\H(a')}K(a,a') \delta(a',\vk) \\ \nn
&\qquad + \int \frac{da'}{a'\H(a')} \left\{ K_1(a,a') \, 
	k^2 \int_{\vq} \delta(a',\vq)\delta(a',\vk-\vq) \right. \\ \nn
&\qquad\qquad\qquad\qquad\quad
	\left. + \, K_2(a,a') k^2 \int_{\vq} \frac{[\vq\cdot(\vk-\vq)]^2}{q^2|\vk-\vq|^2}
	\delta(a',\vq) \delta(a',\vk-\vq) \right. \\ \nn
&\qquad\qquad\qquad\qquad\quad
	\left. +\, K_3(a,a') \int_{\vq} \frac{1}{2} \vq\cdot(\vk-\vq) 
	\lb \frac{\vk\cdot\vq}{q^2} + \frac{\vk\cdot(\vk-\vq)}{|\vk-\vq|^2} \rb 
	\delta(a',\vq)\delta(a',\vk-\vq)   \right. \\
&\qquad\qquad\qquad\qquad\quad
	\left. +\, K_4(a,a') \, k^2
	\lb \frac{\theta(a',\vk)}{-\H(a') f} - \delta(a',\vk) \rb   \right\} \ .
\label{eq:quadeuler}
\end{align}
Notice that when Eqs.~(\ref{eq:deltaexp}) and~(\ref{eq:thetaexp}) are inserted into the last ($K_4$) term, the contents of the square brackets have the same time-dependence at each order.

In principle, from here we could repeat the procedure from App.~\ref{sec:eftreview-linear}, deriving solutions analogous to $\tilde{\delta}^{(n)}$ for each of the four new terms shown above. However, in this work we are only interested in tree-level counterterms (for which each $\delta$ will be evaluated on the linear solution $\delta^{(1)}$) evaluated at $z=0$, for which the time-dependence has no effect. Therefore, we can simply read off the new second-order contributions to $\delta(a=1,\vk)$ from the equation above:
\begin{align}
\nn
\delta(a=1,\vk)_{\rm counterterm} &= - \int_{\vq}  
	\left\{ c_1 k^2 
	+ c_2 \, k^2 \frac{[\vq\cdot(\vk-\vq)]^2}{q^2|\vk-\vq|^2}
	+ c_3 \, \frac{1}{2} \vq\cdot(\vk-\vq) 
		\lb \frac{\vk\cdot\vq}{q^2} + \frac{\vk\cdot(\vk-\vq)}{|\vk-\vq|^2} \rb \right. \\
&\qquad\qquad \left. 
	+\, c_4\, k^2 \! \lb G_2^{\rm (s)}(\vq,\vk-\vq) - F_2^{\rm (s)}(\vq,\vk-\vq) \rb \right\}
	 \delta^{(1)}(\vq) \delta^{(1)}(\vk-\vq)\ .
\label{eq:newquadraticdeltas}
\end{align}
Then, by taking the correlation of each term in Eq.~(\ref{eq:newquadraticdeltas}) with two other linear $\delta$ solutions and symmetrizing over external momenta, we obtain the counterterms listed in Eqs.~(\ref{eq:bc1}). For example, $B_{c_1}$ is obtained from
\begin{align}
(2\pi)^3 \delta_{\rm D}(\vk_1+\vk_2+\vk_3) B_{c_1}(k_1,k_2,k_3) &= 
	\left\la -c_1 k_1^2 \int_{\vq} \delta^{(1)}(\vq) \delta^{(1)}(\vk_1-\vq) 
	\delta^{(1)}(\vk_2) \delta^{(1)}(\vk_3) \right\ra + \text{2 perms}\ .
\end{align}

\section{Divergences and Renormalization in Scaling Universes\label{sec:renormalization}}

In this appendix we want to illustrate that in scale-free universes where the linear power spectrum is given by a power law
\be
P_{11}(k)=\frac{1}{\knl^3}\left(\frac{k}{\knl}\right)^n\ ,
\ee
the one-loop power spectrum and bispectrum present UV divergences. These divergences should, and indeed we show that they can, be cancelled by a suitable chosen combination of the counterterms we have included in the EFT. These are the usual linear $\co$ counterterm, as well as the three new quadratic ones. This result can be interpreted as a sufficient condition that the EFT counterterms are present and induced by the short distance fluctuations. Notice that it is a sufficient but not necessary condition, as even if the counterterms were not needed to cancel any UV divergence, they would still be allowed on symmetry grounds. Furthermore this calculation can be seen as a further check, together with the verification that IR divergences cancel, that our algebra is correct.

Notice that in a scaling universe, the time dependence of the EFT parameters is completely determined by the scaling symmetry present in these universes. The finite part and the UV divergent part, which represents the proper counterterm, as it literally counters a loop diagram, have different time dependences. In particular, the time dependence of the UV divergent part must be exactly the same as the one of the divergent loops. Since in this section we are interested in showing that the UV divergences can be cancelled, we focus only on the UV divergent part of the counterterms. We will consider  only the $n=-1$ case, which has just a logarithmic divergence. Higher $n$'s will have additional subleading divergences that will be very similarly cancelled by higher derivative counterterms that we do not study here.

We start by noticing that the one-loop power spectrum is divergent, the divergence coming from the $P_{13}$ diagram. This requires a renormalization by using the $\co$ counterterm, as described in general in~\cite{Carrasco:2012cv} and for scaling universes in~\cite{Pajer:2013jj,Carrasco:2013sva}. We obtain
\beq
\bar c_1 = - \frac{122\pi}{315} \frac{1}{\knl^2}\log\!\lp \frac{\Lambda}{k_{\rm min}} \rp\ ,
\eeq
where as in (\ref{eq:cbartocs}) we have that $\bar c_1=\frac{2\pi}{\knl^2}\co$. Here we take the parameter $\zeta=2$ as we are interested in canceling the UV divergent part of the one-loop diagrams.

Next we move to the Bispectrum. The divergent part is given by:
\bea
&& B_\text{1-loop,UV}= -\frac{\pi }{169785
   k_1^3 k_2^3 k_3^3 \knl^6} \log\left(\frac{\Lambda }{\knl}\right)\; \times \\ \nonumber
&&   \quad   \left(12409 k_1^9+29479 k_1^7
   \left(k_2^2+k_3^2\right)-11461 k_1^6
   \left(k_2^3+k_3^3\right)+k_1^5 \left(-30427 k_2^4+104866
   k_2^2 k_3^2-30427 k_3^4\right)\right.\\ \nonumber
&&\quad \left.   +k_1^4 \left(-30427
   k_2^5+11461 k_2^3 k_3^2+11461 k_2^2 k_3^3-30427
   k_3^5\right)-11461 k_1^3 \left(k_2^2-k_3^2\right)^2
   \left(k_2^2+k_3^2\right)\right. \\ \nonumber
&&  \quad \left.+k_1^2 \left(29479 k_2^7 
   +104866
   k_2^5 k_3^2+11461 k_2^4 k_3^3+11461 k_2^3
   k_3^4+104866 k_2^2 k_3^5+29479
   k_3^7\right)+  \right. \\ \nonumber
   && \quad \left. (k_2-k_3)^2 (k_2+k_3)\right.\\ \nonumber
   &&\quad\qquad\left. \left.\times\left(12409
   k_2^6+12409 k_2^5 k_3+54297 k_2^4 k_3^2\right.+42836 k_2^3
   k_3^3+54297 k_2^2 k_3^4+12409 k_2 k_3^5+12409
   k_3^6\right)\right)   .
   \eea

Simple algebra show that the divergent part can be written as a linear combination of our three quadratic counterterms in~(\ref{eq:bc1}) plus the $\co$ counterterm in~(\ref{eq:speedofsoundterm}), all evaluated with $\zeta=2$. We obtain
\beq
-B_\text{1-loop, \rm UV}(k_1,k_2,k_3)= c_1 \, B_{c_1} (k_1,k_2,k_3)+
	  c_2  B_{c_2}(k_1,k_2,k_3)+ c_3 B_{c_3} (k_1,k_2,k_3)\ .
\eeq
where
\beq
c_1 =\frac{892\pi}{56595 }\frac{1}{\knl^2} \log\!\lp \frac{\Lambda}{k_{\rm min}} \rp, \quad
	c_2 = \frac{2914\pi}{33957 }\frac{1}{\knl^2} \log\!\lp \frac{\Lambda}{k_{\rm min}} \rp, \quad
	c_3 = \frac{27998\pi}{52595 }\frac{1}{\knl^2} \log\!\lp \frac{\Lambda}{k_{\rm min}}  \rp\ .
\eeq
This completes our demonstration that the UV divergences can be reabsorbed.

\section{A Simplification to the IR Resummation \label{app:IRresum}}

In~\cite{Senatore:2014via} it was shown that the IR-resummed expression for the dark matter power spectrum up to order~$N$ in~$\epsilon_{\delta<}$ was given by a convolution integral of the following form
\be\label{eq:delta7} 
\left.P_{\delta\delta}(k;t_1,t_2)\right|_N=\sum_{j=0}^N \int \frac{d^3k'}{(2\pi)^3}\; M_{||_{N-j}}( k, k';t_1,t_2)\; P_{\delta\delta,\,j}(k';t_1,t_2)\ ,
\ee
where $P_{\delta\delta,\,j}(k';t_1,t_2)$ is the power spectrum evaluated to order $j$ expanding both in $\epsilon_{\delta<}$ and $\epsilon_{s<}$.
 Here the convolution matrix $M$ is given by
\be\label{eq:Meq}
M_{||_{N-j}}( k, k';t_1,t_2)=\int d^3 r\; d^3 q \; P_{||_{N-j}}(\vec r|\vec q;t_1,t_2) \; e^{i \vec k \cdot\vec r} \;e^{-i \vec k' \cdot\vec q}\ ,
\ee
with $P_{||_{N-j}}(\vec r|\vec q;t_1,t_2)$ is related to the probability for a particle of starting at position $\vec q$ and ending up at position $\vec r$:
\be\label{eq:Peq}
P_{||_{N-j}}(\vec r|\vec q;t_1,t_2)=\int\frac{d^3 k''}{(2\pi)^3}\; e^{-i \vec k''\cdot (\vec q-\vec r)}\;F_{||_{N-j}}(\vec q,\vec k'';t_1,t_2)\ ,
\ee
where $F_{||_{N-j}}$ is defined in eq.~(43) of~\cite{Senatore:2014via}. In~\cite{Senatore:2014via}, $M$ was evaluated from (\ref{eq:Meq}) for every $k$ and $k' $ by performing a bi-dimensional Fast Fourier Transform (FFT). It is however possible to alternatively evaluate $M$ by performing, for each $k$ of interest, a one-dimensional FFT. This can be seen by plugging $(\ref{eq:Peq})$ into~(\ref{eq:Meq}). The $\vec r$ integral can be done leading to a $(2\pi)^3\delta_{\rm D}(\vec k''+\vec k)$, which in turns allows us to evaluate the $\vec{k}''$ integral. We then obtain
\be\label{eq:Meq2}
M_{||_{N-j}}( k, k';t_1,t_2)=\int d^3 q \; F_{||_{N-j}}(\vec q,-\vec k;t_1,t_2) \; e^{i (\vec k-\vec k') \cdot\vec q}\ .
\ee
Plugging~(\ref{eq:Meq2}) into~(\ref{eq:delta7}), we can then perform then angular $\vec{k}'$ integrals analytically, leaving
\be
\label{eq:pddnew}
\left.P_{\delta\delta}(k;t_1,t_2)\right|_N 
	=\sum_{j=0}^N \int dk' \; \hat{M}_{||_{N-j}}(k,k';t_1,t_2)\; P_{\delta\delta,\,j}(k';t_1,t_2)\ ,
\ee
where
\be
\label{eq:mhat}
\hat{M}_{||_{N-j}}(k,k';t_1,t_2)
	= \frac{1}{2\pi^2} \int d^3q \, \frac{k' \sin(k'q)}{q} \;e^{i\vec{k}\cdot\vec{q} }\;
	F_{||_{N-j}}(\vec q,\vec k;t_1,t_2)\ ,
\ee
and where we used that $F_{||_{N-j}}(\vec q,-\vec k;t_1,t_2)=F_{||_{N-j}}(\vec q,\vec k;t_1,t_2)$.
Specifically, the $F_{||_{N-j}}$ functions that are needed for the equal-time two-loop power spectrum computation are given by
\begin{align}
F_{||_0}(\vec{q},\vec{k};t) &= \exp\! \lb -\frac{1}{2} A_{ij,1}(\vec{q};t) k^i k^j \rb\ , \\
F_{||_1}(\vec{q},\vec{k};t) &= F_{||_0}(\vec{q},\vec{k};t) \lb 1
	+  \frac{1}{2} A_{ij,1}(\vec{q};t) k^i k^j  \rb\ , \\
F_{||_2}(\vec{q},\vec{k};t) &= F_{||_0}(\vec{q},\vec{k};t) \lb 1 
	+ \frac{1}{2} A_{ij,1}(\vec{q};t) k^i k^j  
	+ \frac{1}{8} \lb A_{ij,1}(\vec{q};t) k^i k^j \rb^2  \rb\ ,
\end{align}
where
\beq
A_{ij,1}(\vec{q};t) = X(q;t)_1\, \delta_{ij} + Y(q;t)_1\, \hat{q}_i \hat{q}_j
\eeq
and
\begin{align}
X(q;t)_1 &= \frac{1}{2\pi^2} \int_0^\infty dk \, \exp \lb -\frac{k^2}{\Lambda_{\rm resum}^2} \rb
	P_{11}(k;t) \lb \frac{2}{3} - 2 \frac{j_1(kq)}{kq} \rb\ ,\\
Y(q;t)_1 &= \frac{1}{2\pi^2} \int_0^\infty dk \, \exp \lb -\frac{k^2}{\Lambda_{\rm resum}^2} \rb
	P_{11}(k;t) \lb -2j_0(kq) + 6 \frac{j_1(kq)}{kq} \rb\ ,
\end{align}
given in terms of the spherical Bessel functions $j_i(x)$.
The angular $\vec{q}$ integrals in Eq.~(\ref{eq:mhat}) can also be performed analytically, leaving a single~$q$ integral in that can then be evaluated via an FFT at each desired~$k$ value. After this, the remaining $k'$ integral in~(\ref{eq:pddnew}) is trivial to carry out numerically. In this work, as in~\cite{Senatore:2014via}, we use $\Lambda_{\rm resum} = 0.1\invMpc$.

It is also possible to re-derive this simplification working from the derivation presented~\cite{Senatore:2014via}. We can start from eq.~(44) of~\cite{Senatore:2014via}, that reads:
\be\label{eq:delta3} 
\left.P_{\delta\delta}(k_1;t_1,t_2) \right|_{N} \\ 
	=\int d^3 q \  e^{- i \vec k_1 \cdot \vec q}\;  
	\sum_{j=0}^N\left[ F_{||_{N-j}}(\vec q,\vec k_1;t_1,t_2)\cdot 
	K(\vec q,\vec k_1;t_1,t_2)_{j}\right]\ ,
\ee
where $K(\vec q,\vec k_1;t_1,t_2)_{j}$ is such that the power spectrum to order  $j$ expanding both in $\epsilon_{\delta<}$ and $\epsilon_{s<}$ is given by
\be
P_{\delta\delta}(k;t_1,t_2)_{j}=\int d^3 q \; e^{i \vec k\cdot\vec q}\;K(\vec q,-\vec k;t_1,t_2)_j\ .
\ee
It is  useful to manipulate the above expression by multiplying by 1 written as
\be
1=\int \frac{d^3 k_1'}{(2\pi)^3}\; (2\pi)^3\delta_{\rm D}(\vec k'_1-\vec k_1)
=\int\frac{d^3 k_1'}{(2\pi)^3} \int d^3 q' \; e^{i\, \vec q'\cdot(\vec k'_1-\vec k_1)} \ ,
\ee
so that, by exchanging $\vec k_1$ with $\vec k'_1$ when useful, we can write
\bea\label{eq:delta4} 
&&\left.P_{\delta\delta}(k_1;t_1,t_2) \rangle\right|_{N} \\  \nonumber
&& \qquad =\sum_{j=0}^N \int \frac{d^3 k'_1}{(2\pi)^3} \left[ \int d^3 q'\;  e^{i\cdot(\vec k_1'- \vec k_1) \cdot \vec q'}\;   F_{||_{N-j}}(\vec q,\vec k_1;t_1,t_2)\right]
\cdot \left[\int d^3 q\; e^{- i \vec k_1' \cdot \vec q} K(\vec q,\vec k_1';t_1,t_2)_{j}\right]\\   \nonumber 
\eea
Now, as done in~\cite{Senatore:2014via}, we perform the approximation to replace $F_{||_{N-j}}(\vec q,\vec k_1;t_1,t_2)\to F_{||_{N-j}}(\vec q',\vec k_1;t_1,t_2)$, as this amounts to doing a mistake proportional to the gradient of the IR displacements. We therefore obtain directly Eq.~(\ref{eq:delta7}), but with the matrix $M$ directly in the form~(\ref{eq:Meq2}).

\section{Consistency Check using Other Simulations}
\label{sec:othersims}

As a check on our primary results presented in Sec.~\ref{sec:results}, we have also performed the same procedure on a second, independent set of simulations. These simulations are based on three realizations with $L_{\rm box}=1600 h^{-1}$Mpc and $512^3$ grid points, assuming  a flat $\Lambda$CDM model with $\Omega_{\rm b}h^2=0.0226$, $\Omega_{\rm c}h^2=0.11$, $\Omega_\Lambda=0.734$, $h=0.71$, $\Delta_\mathcal{R}^2(k_0)=2.43\times 10^{-9}$, and $n_{\rm s}(k_0)=0.963$, where $k_0=0.002{\rm Mpc}^{-1}$; more details can be found in~\cite{Schmittfull2013}. These cosmological parameters give, over the range $0.25\hinvMpc \lesssim k \lesssim 0.60 \hinvMpc$, $P_{11}(k) \approx \frac{(2\pi)^3}{\knl^3} \lp \frac{k}{\knl} \rp^n$ with $n\approx-2.1$ and $\knl\approx 4.45 \hinvMpc$. 

The matter bispectrum we compare to was reconstructed from these simulations using an expansion of separable basis functions (see~\cite{Schmittfull2013}), rather than the more traditional methods described in Sec.~\ref{sec:simulations}, and for this reason, the error estimate in Eq.~(\ref{eq:biserrcv}) cannot be straightforwardly applied. Instead, we expect there to be two main contributions to the errorbars: scatter between the three realizations, which can be used as a rough estimate for the cosmic variance, and the fact that we truncate the series of basis functions after a finite number of terms, which introduces a separate systematic error. We estimate the uncertainty incurred by this truncation---specifically, the difference between the exact $B_\text{nonlinear}$ and the $B_\text{nonlinear}$ reconstructed from the basis functions---by calculating this difference for $B_\text{tree}$, for which we know the exact function. We then estimate the total uncertainty in our measurements of $B_\text{nonlinear}$ by adding this truncation uncertainty, double the scatter between realizations (we double the measured scatter because three realizations will generally not correctly estimate the cosmic variance, and so we prefer to overestimate it), and an extra 2\% systematic error. This is a rather gross estimate of the errorbars, but is still useful as a visual aide to determine whether the theoretical predictions are a reasonable match to the data. However, different triangles will have errors that are highly correlated, since the error on each triangle will arise from combined errors on the measured coefficients of each function in the separable basis, and the reader should keep in mind that we ignore these correlations in our fits and plots.

\begin{figure}[t]
\begin{center}
\includegraphics[width=0.95\textwidth]{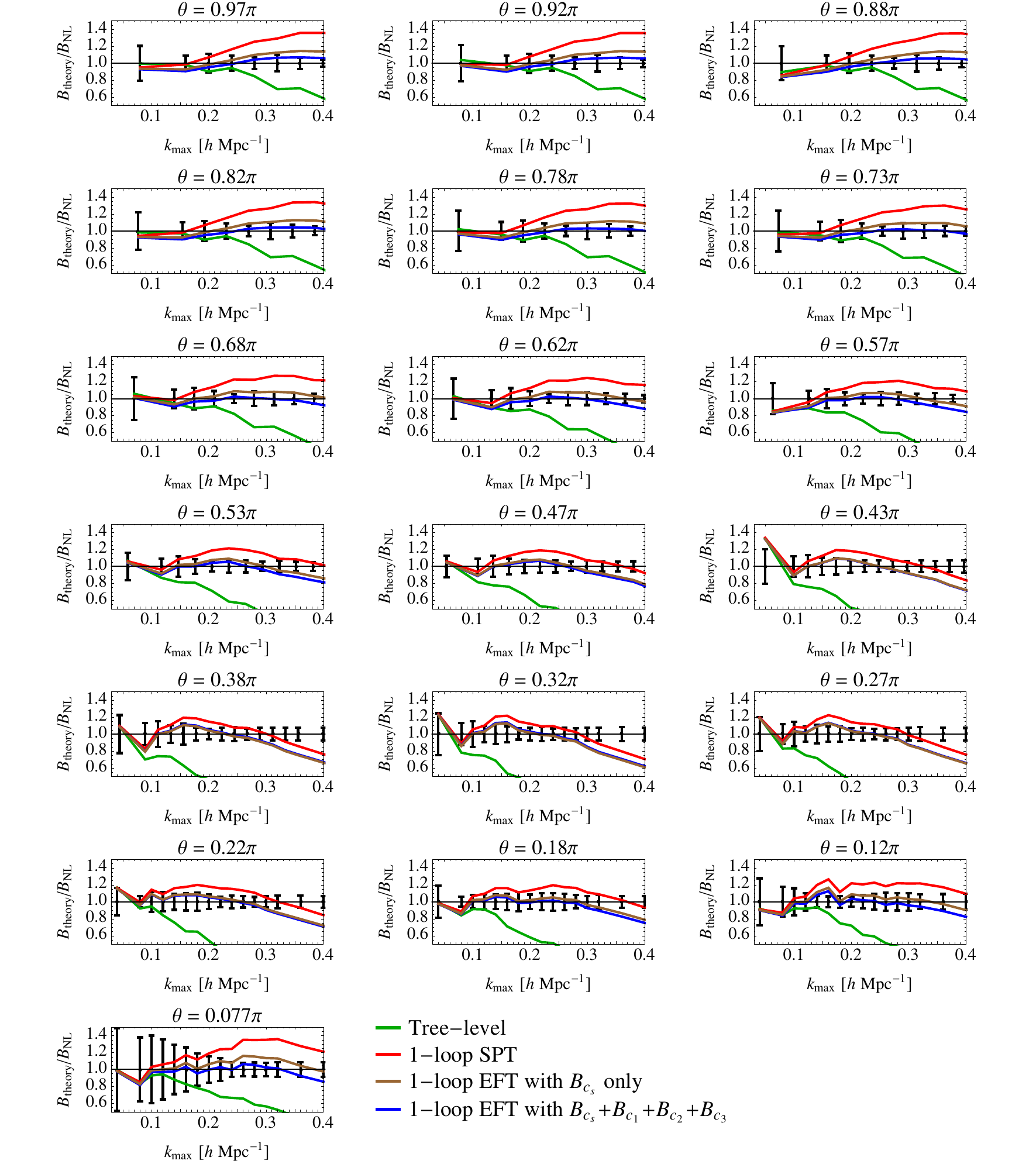}
\caption{ \label{fig:othersims-fits-isos} \footnotesize\it  Comparison of various theory curves, normalized to nonlinear bispectrum data for isosceles triangles ($k_2=k_1$) measured from the simulations in App.~\ref{sec:othersims}. The blue curve is fit using all triangles with maximum side length less than $0.3\hinvMpc$. It is clear that the EFT predictions significantly improve upon the tree-level and SPT predictions, particularly for squeezed and flat configurations. }
\end{center}
\end{figure}

\begin{figure}[t]
\begin{center}
\includegraphics[width=0.95\textwidth]{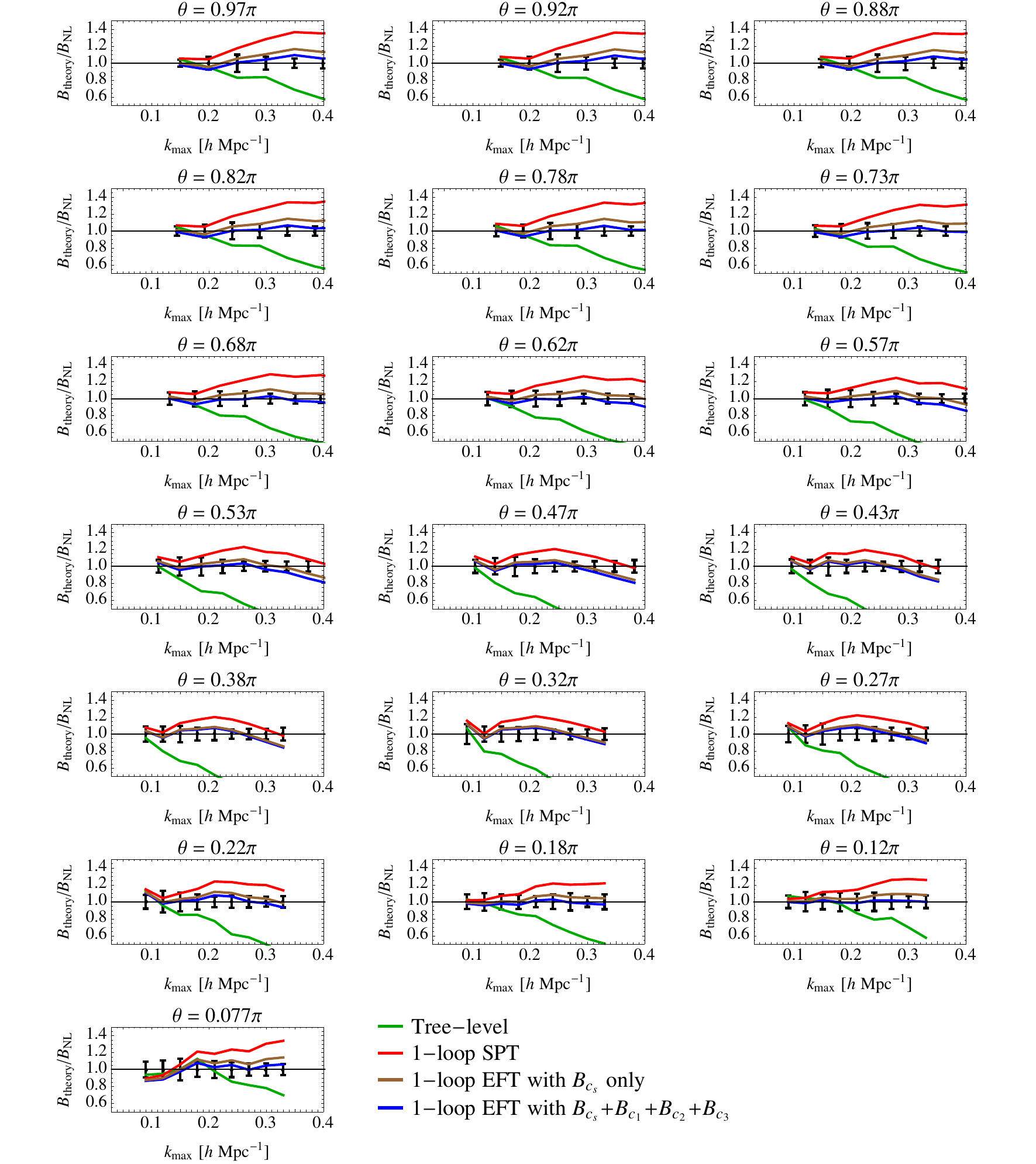}
\caption{ \label{fig:othersims-fits-3o2} \footnotesize\it  As Fig.~\ref{fig:othersims-fits-isos}, but for triangles with $k_2=1.5k_1$.}
\end{center}
\end{figure}

\begin{figure}[t]
\begin{center}
\includegraphics[width=0.95\textwidth]{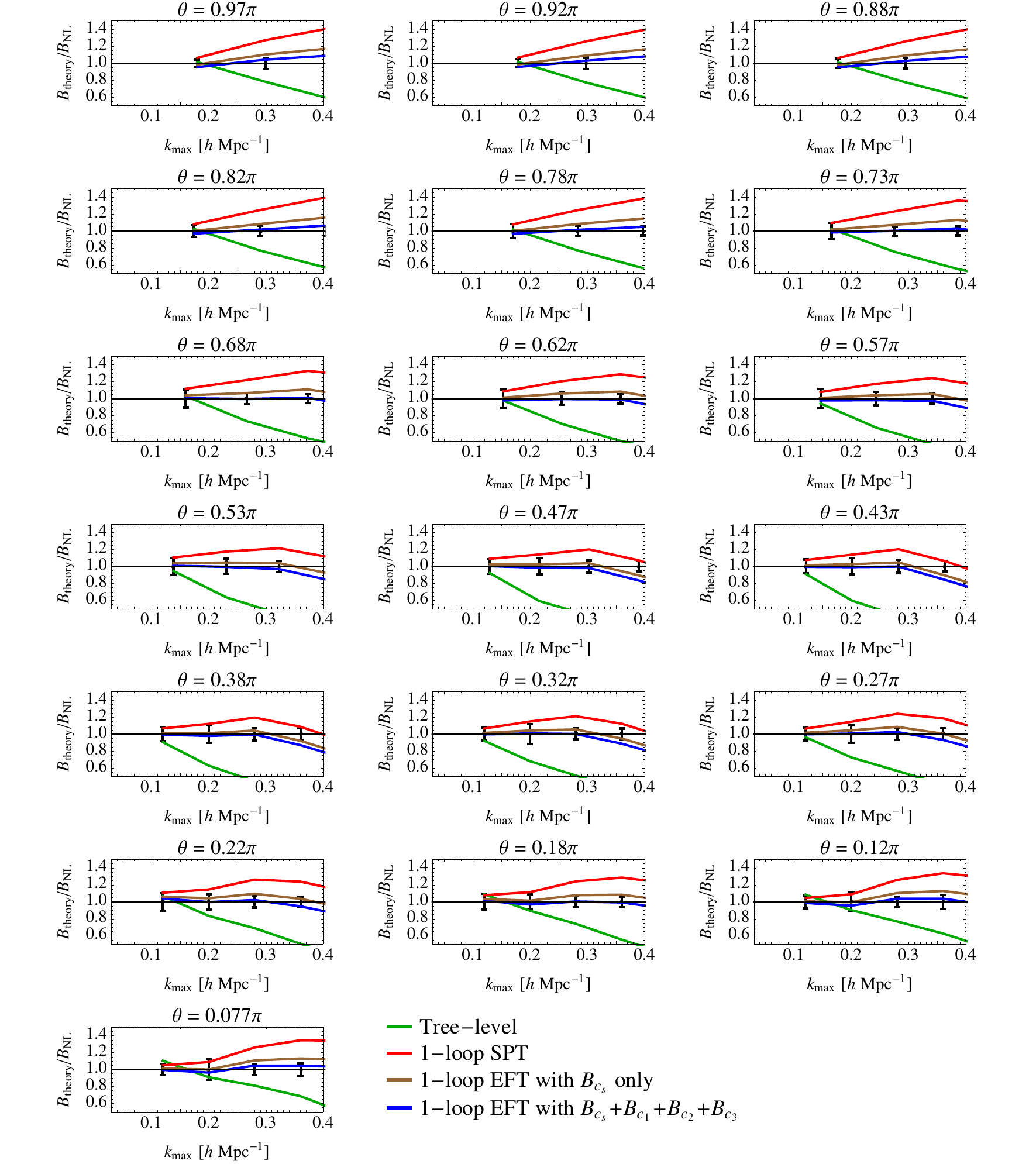}
\caption{ \label{fig:othersims-fits-2} \footnotesize\it  As Fig.~\ref{fig:othersims-fits-isos}, but for triangles with $k_2=2k_1$.}
\end{center}
\end{figure}

From there, we follow the same procedure as in Secs.~\ref{sec:csfit} and~\ref{sec:results}. Fitting the one-loop EFT prediction to the matter power spectrum yields 
\beq  
\co =( 1.58 \pm {  0.16}) \times \frac{1}{2\pi} \lp \frac{\knl}{\invMpc} \rp^{2} \qquad (\text{1-$\sigma$}) , 
\eeq
and we use this value, plus values of $c_1$, $c_2$, and $c_3$ obtained by fitting the one-loop bispectrum formula to all bispectrum triangles with maximum side length less than $0.3\hinvMpc$, to obtain the results shown in Figures~\ref{fig:othersims-fits-isos},~\ref{fig:othersims-fits-3o2}, and~\ref{fig:othersims-fits-2}. Once again, the improvement of the EFT prediction over one-loop SPT is evident. The fact that such improvement occurs for two separate sets of simulations demonstrates that the improvement arises from properties of the EFTofLSS, rather than the details of any particular simulation.

\end{document}